\documentclass[prd,twocolumn,nofootinbib,superscriptaddress]{revtex4-2}

\usepackage{preamble}
\usepackage{orcidlink}

\begin{document}
\title{The ringdown of a black hole surrounded by a thin shell of matter}
\author{Andrew Laeuger\, \orcidlink{0000-0002-8212-6496}}
\email{alaeuger@caltech.edu}
\affiliation{California Institute of Technology, Pasadena, CA 91126}
\author{Colin Weller\, \orcidlink{0000-0001-5173-5638}}
\affiliation{California Institute of Technology, Pasadena, CA 91126}
\author{Dongjun Li\, \orcidlink{0000-0002-1962-680X}}
\affiliation{Illinois  Center  for  Advanced  Studies  of  the  Universe \&
Department of Physics, University of Illinois at Urbana-Champaign, Urbana, Illinois 61801, USA}
\author{Yanbei Chen\, \orcidlink{0000-0002-9730-9463}}
\affiliation{California Institute of Technology, Pasadena, CA 91126}
\date{\today}

\begin{abstract}
    Recent studies have shown that far-field perturbations to the curvature potential of a black hole spacetime may destabilize its quasinormal mode (QNM) spectrum while only mildly affecting time-domain ringdown signals. In this work, we study the QNM spectrum and ringdown behavior of a Schwarzschild black hole with a far-field perturbation to its physical environment -- a thin matter shell with finite surface tension. After accounting for the dynamics of the interaction between GWs and the shell, we find that the fundamental mode can migrate perturbatively or be destabilized by the appearance of new modes with no analogue in the vacuum case, much like studies of ``bumps" in the curvature potential. However, unlike these previous works, we find that the coupling between metric perturbations and oscillations of the shell also sources weakly-damped QNMs which are exclusive to the polar sector. We then study whether the analysis tools of least-squares QNM fits and the full and rational ringdown filters can clearly identify the signatures of the shell in representative ringdown waveforms. We conclude that ringdown at sufficiently early times is insensitive to the shell; weakly-damped QNMs (in the polar sector) and echoes, which may enable the analysis methods considered here to infer the presence of a shell, only appear at late times and are generally weak. 
\end{abstract}

\maketitle

\section{Introduction}
Throughout the last decade, general relativity (GR) has withstood tests from dozens of gravitational-wave (GW) detections by the LIGO/Virgo/KAGRA (LVK) Collaboration \cite{LIGOScientific:2016aoc, LIGOScientific:2017vwq, LIGOScientific:2018dkp,LIGOScientificTGR}, the first image of the supermassive black hole (BH) in the core of the massive elliptical galaxy M87 \cite{EventHorizonTelescope:2023gtd}, and the detection of the stochastic GW background \cite{NANOGrav:2023gor}. Black hole ringdown, a phenomenon observed at the conclusion of compact object mergers as the remnant BH relaxes to its equilbrium state, offers another avenue for testing GR with GW observations \cite{IsiTGR,BertiTGR,SilvaTGR,BertiTGR2,DreyerTGR}.

Within the framework of GR, the ringdown of a remnant BH is dominated by quasinormal modes (QNMs) and generally well-described by linear perturbation theory \cite{TeukolskyEquations,Teukolsky:1974yv,Press:1973zz,ZerilliEquation,ReggeWheelerEquation,Moncrief:1974am}. The strain $h\equiv h_++ih_\times$ is given by 
\begin{align}
    h(t)\equiv\sum_{\ell,m,n}A_{\ell m n}e^{-i\omega_{\ell m n} t + i m \phi}{}_{-2}S_{\ell m}(a \omega_{\ell m n};\theta),
\end{align}
where $a = J/M$ is the ratio of the angular momentum $J$ and mass $M$ of the BH. The indices $\ell,m$ label the angular harmonic, $n$ labels the overtone, $\omega_{\ell m n}$ are the linear QNM frequencies -- entirely determined by the BH mass, spin, and electric charge \cite{VishveshwaraQNMs,LeaversMethod,PressQNMs,DetweilerQNMs} -- $A_{\ell m n}$ are the complex-valued amplitudes, and ${}_{-2}S_{\ell m}(\theta)$ are the angular eigenfunctions of the Teukolsky equation \cite{TeukolskyEquations}. This prediction has been confirmed by numerical relativity simulations of black hole mergers in vacuum  \cite{MitmanNRRingdown,GieslerNRRingdown,CheungNRRingdown}, though as of yet, drawing confident conclusions regarding the validity of the ringdown model from compact merger data has proven to be a difficult and contentious pursuit \cite{Isi150914,Cotesta150914,Isi150914comment,Correia150914,Finch150914,Wang150914,Siegel190521}.

Since the initial discovery of QNMs, many different methods have been employed to calculate QNM frequencies both in GR and its effective field theory extensions, such as WKB approximations \cite{Yang:2012he,Iyer:1986np,Schutz:1985km}, inverse-potential estimations \cite{Ferrari:1984zz}, numerical shooting, continued fraction methods \cite{LeaversMethod,Chandrasekhar:1975zza, Mark:2014aja}, and eigenvalue perturbation methods \cite{Li:2022pcy, Hussain:2022ins, Wagle:2023fwl, Zimmerman:2014aha, Mark:2014aja, WellerBumpyBHs}. As the BH spectroscopy program continues to develop, greater attention has been paid to the question of whether the astrophysical environment surrounding a BH will affect tests of GR in ringdown signals \cite{BarausseEnvironmentalEffects,BarausseEnvironmentalEffects2,Konoplya:2019sns,Torres:2022fyf,Kavanagh:2020cfn, Konoplya:2018yrp, Brito:2023pyl,Antonini:2016gqe}. Furthermore, exotic environments emerging from beyond-Standard-Model physics may also alter the ringdown signature from binary BH coalescences, and thus have also received recent attention \cite{Kiuchi:2011re,Medved:2003rga,Konoplya:2019sns,Brito:2023pyl,Spieksma:2024voy,PezzellaHaloQNMs,SpeeneyDMSpikes,SpeeneyPolarPerturbations,PezzellaHaloQNMs,CardosoHaloPerturbationFormulas}. 

A number of recent works have found that when a small ``bump" (a toy model for some localized physical feature) is placed far from the BH in the curvature potential appearing in the perturbation theory wave equation, the QNM spectrum is destabilized -- that is, the frequencies migrate an amount far larger than the characteristic scale of the bump \cite{CheungFundamentalInstability,CardosoQNMInstabilitySignificance,IanniccariInstability,CourtyInstability,JaramilloPseudospectrum,CaoQuantumCorrectedInstability,CowndenInstability,Daghigh:2020jyk}. The destabilized spectrum could challenge the precision of the BH spectroscopy program which aims to infer the mass, spin, and electric charge of the remnant BH from measuring QNMs \cite{Israel:1967wq,VishveshwaraQNMs,LeaversMethod,PressQNMs,DetweilerQNMs,IsiRingdownPE,Bhagwat:2019dtm,SiegelRingdownPE,BertiRingdownPE,WangRingdownPE}. Fortunately, further works examining time-domain ringdown gravitational waveforms produced in such systems have found evidence that the ringdown waveforms are stably perturbed, and thus the BH spectroscopy program is not completely jeopardized \cite{BertiTimeDomainStability,JaramilloInstabilitySignatures,BarausseEnvironmentalEffects,BarausseEnvironmentalEffects2,OshitaInstability}. Still others have considered using BH greybody factors \cite{OshitaGreybodyI,OshitaGreybodyII,OshitaGreybodyStability,PaniGreybodyStability,Konoplya:2025ixm} or the ringdown filters \cite{SizhengFilters,SizhengFilterOvertone,SizhengFilterCleaning,LuFilterStatistics} (both of which are derived from the same function of frequency) as an alternative tool for parameterizing ringdown signals, and found that it, unlike the QNM spectrum, exhibits stability under similar perturbations to the potential in the wave equation. 

In this paper, we investigate the QNM spectrum and ringdown behavior of a Schwarzschild BH surrounded by a simple matter distribution: an infinitely thin shell of matter with finite surface tension. Instead of directly modifying the wave equation potential which emerges from a linear expansion of Einstein's equations on a Schwarzschild background, we complete a full treatment of the behavior of metric perturbations in this Schwarzschild with shell spacetime, including the interaction of the metric perturbations with the matter in the shell. As we will see later on, our inclusion of the interaction physics directly results in new ringdown features which have not been identified in previous works examining the effect of localized exterior perturbations on QNMs. A few previous studies \cite{NaidooGWShells,BishopGWShells,Acuna-CardenasGWShells,Konoplya:2018yrp} have also considered the propagation of GWs in environments surrounded by shells of matter; however, they consider different background spacetimes or apply different mathematical procedures and in general do not investigate the ringdown behavior to the extent done here. Note that in this work, we will often use the phrase ``vacuum case" to describe the system where the shell of matter is absent.

The remainder of this paper proceeds as follows. In Sec. \ref{sec: mathematical construction}, we describe how the existence of a thin matter shell modifies the propagation of metric perturbations on the entire radial domain outside the horizon. In Sec. \ref{sec: qnms}, we examine the three types of QNMs which can be sourced by this system and provide a physical intuition for the impact of the shell parameters on the location of their frequencies frequencies in the complex plane. 
In Sec. \ref{sec: ringdown fits}, we turn our attention to a radial infall waveform in this spacetime and examine how well the vacuum and shell system QNM frequencies fit the ringdown portion. In Sec. \ref{sec: Ain on real axis}, we study the effect of the shell on the quantity $A_\text{in}$, which contains information regarding the propagation of gravitational waves in this spacetime. In Sec. \ref{sec: ringdown filters}, we consider the ringdown filters proposed in \cite{SizhengFilters} -- constructed from $A_\text{in}$ -- and compare the efficacy of the vacuum and shell system filters in filtering out the ringdown component of the shell system radial infall waveform. Finally, in Sec. \ref{sec: conclusion}, we conclude and discuss directions in which to extend this program.

In this work, we employ geometrized units $G=c=1$, and in our computations of QNM frequencies and ringdown waveforms, we follow the convention of \cite{LeaversMethod} where the black hole mass is $M=1/2$. All data is publicly available at \cite{alaeuger_2025_15586889}.

\section{Mathematical Construction}
\label{sec: mathematical construction}
In this section, we outline a mathematical procedure for computing how metric perturbations propagate in a spacetime consisting of a Schwarzschilld black hole surrounded by a thin spherical shell of matter. We then describe how this procedure can be applied to extract quantities relevant for understanding the ringdown of such a spacetime.

\subsection{Wave Propagation in Vacuum}
For a background metric of the form
\begin{equation}
    ds^2=-f(r)^2dt^2+\frac{1}{h(r)}dr^2+r^2d\Omega^2,
\end{equation}
small perturbations to the metric $\delta g_{\mu\nu}$ are commonly written in the Regge-Wheeler gauge (RWG), given by 
\begin{equation}
    \delta g_{\mu\nu}=\begin{pmatrix}
        \vspace{0.1cm}
        fH_0 & H_1 & -h_0\frac{1}{\sin\theta}\partial_\phi  & h_0\sin\theta\partial_\theta 
        \\
        \vspace{0.1cm}
        * & \frac{1}{h}H_2 & -h_1\frac{1}{\sin\theta}\partial_\phi  & h_1\sin\theta\partial_\theta 
        \\
        * & * & r^2K & 0
        \\
        * & * & * & r^2\sin^2\theta K
    \end{pmatrix}
    Y_{\ell m},
\end{equation}
where $Y_{\ell m}(\theta,\phi)$ are the spherical harmonics \cite{MaggioreVol2}. Summation over $\ell$ and $m$, where $|m|\leq\ell$, is implied. Components labeled with $*$ can be extracted via the symmetry of $\delta g_{\mu\nu}$, and for brevity, we have suppressed the $(t,r)$ dependence and mode indices $\ell$ and $m$ in the RWG perturbation variables $H_0$, $H_1$, $H_2$, $K$, $h_0$, and $h_1$. For the remainder of this work, we will continue to suppress the mode indices, with $\ell=2$ always implied\footnote{We have checked that for $\ell=3$, the analogous results to those presented here are qualitatively similar.}. The linearized Einstein equations require that $H_0=H_2\equiv H$. 

The RWG metric perturbation can be decomposed into polar components -- those featuring $H$, $H_1$, and $K$ -- and axial components -- those featuring $h_0$ and $h_1$. The polar sector perturbation variables can be combined into the Zerilli master function $Z$, given in the frequency domain (assuming time evolution $\propto e^{-i\omega t}$) by
\begin{equation}
    Z(\omega,r)=\frac{r^2}{\lambda r+3M}K(\omega,r)+\frac{rf_S(r)}{i\omega (\lambda r+3M)}H_1(\omega,r),
\end{equation}
where $f_{\text{S}}(M,r)\equiv 1-2M/r$ is the canonical form of $f$ in Schwarzschild and $\lambda\equiv(\ell-1)(\ell+2)/2$. The Zerilli master function satisfies a wave equation
\cite{ZerilliEquation}:
\begin{equation}
    \frac{\partial^2}{\partial r_*^2}Z(\omega,r)+(\omega^2-V_{\text{Z}}(M,r))Z(\omega,r)=S(\omega,r),
\end{equation}
where $r_*(r)\equiv r+2M\ln(r/2M-1)$, $S(\omega,r)$ is a source term, and $\omega$ is defined with respect to the coordinate time of the interior metric $t$. Throughout this work, we will use both $r$ and $r_*$ to describe radial coordinates but suppress the dependence on the other coordinate in the pair, i.e. writing $r(r_*)$ in functions defined in terms of $r_*$ as just $r$, and vice versa. The Zerilli potential is
\begin{multline}
    V_{\text{Z}}(M,r)\equiv \frac{f_{\text{S}}(M,r)}{r^3(\lambda r+3M)^2}
    \\
    \times(2\lambda^2(\lambda+1)r^3+6\lambda^2Mr^2+18\lambda M^2r+18M^3).
\end{multline} 
We denote the homogeneous solutions to the Zerilli equation by $Z$ and the solutions with a source by $\mathcal{Z}$.

In the axial sector, the perturbation variables are related to one another in vacuum by
\begin{equation}
    h_0(\omega,r)=\frac{f_{\text{S}}(M,r)}{i\omega}\frac{\partial}{\partial r}(f_{\text{S}}(M,r)h_1(\omega,r)),
    \label{eq: axial perturbation conversions}
\end{equation}
and the Regge-Wheeler function, given by 
\begin{equation}
    Q(\omega,r)\equiv \frac1rf_{\text{S}}(M,r)h_1(\omega,r),
    \label{eq: RW function}
\end{equation}
satisfies the Regge-Wheeler equation \cite{ReggeWheelerEquation}, an analogous wave equation with potential
\begin{equation}
    V_{\text{RW}}(r)=f_{\text{S}}(M,r)\Bigr(\frac{\ell(\ell+1)}{r^2}-\frac{6M}{r^3}\Bigr).
\end{equation}
In a pure vacuum spacetime, one can simply integrate the master functions through space using the Zerilli and Regge-Wheeler equations, and from the master functions evaluated on the entire domain, ultimately reconstruct $\delta g_{\mu\nu}$. However, the thin shell introduces a region where the spacetime is no longer vacuum, complicating the picture. We now outline our procedure for including the stress-energy of the shell into the propagation of metric perturbations.

\subsection{Imposing the Thin Shell} 
\label{sec: junction conditions}
In this work, we aim to further the understanding of how localized perturbations to a Schwarzschild spacetime impact the ringdown of such a system. However, instead of applying such a perturbation directly to the wave equation potentials, as many previous studies have done, we wish to impose the perturbation upon the astrophysical environment itself. A thin spherical shell of matter surrounding the Schwarzschild BH offers a simple toy model for astrophysical features, such as gas clouds, which may surround the compact object mergers which generate our observed GW signals.

Our model of this thin spherical shell follows that of \cite{PaniGravastar}. Namely, we characterize the shell by the surface stress-energy tensor for a perfect fluid. When unperturbed, it appears as 
\begin{equation}
    S_{jk}=(\Sigma-\Theta)u_ju_k-\Theta\gamma_{jk}.
    \label{eq: static stress energy}
\end{equation}
The indices $j$ and $k$ run over $t$, $\theta$, and $\phi$, $u^\alpha\propto\partial_t$ is the four-velocity of the matter, $\gamma_{jk}\equiv g_{jk}-n_jn_k$ is the induced metric on the shell with $n_i$ being the unit normal vector to the shell, $\Sigma$ is the surface energy density of the shell, and $\Theta$ is the surface tension.

A number of previous studies (e.g., \cite{FrauendienerShell,LeMaitreStability,BradyStability,GoncalvesShells,KijowskiStability,LoboStability,HuanMembranes}) have carried out in-depth explorations of the mechanical stability of static shells surrounding black holes and found that a broad range of reasonable equations of state can admit thin matter shells which are statically and dynamically stable as long as their radius is not too close to the horizon. The mechanical stability of this shell is not the central focus of this work, so for our purposes here, we just apply a simple equation of state $\delta\Theta= -v_\text{s}^2\delta\Sigma$, with $v_{\text{s}}$ being the speed of sound in the shell matter, and assume the system is stable in the static case.

To account for the stress-energy of the shell, we rely on Israel's junction conditions \cite{IsraelJuncConds}, which relate the discontinuities in the extrinsic curvature tensor $K_{jk}$ over the boundary of a thin surface of matter to its surface stress-energy tensor. Writing the discontinuities over the shell in quantity $a$ as
\begin{equation}
    [[a]]\equiv\Bigr(\lim_{r\rightarrow R^+}a\Bigr)-\Bigr(\lim_{r\rightarrow R^-}a\Bigr),
\end{equation}
Israel's junction conditions can be expressed as
\begin{equation}
    [[K_{jk}]]=8\pi(S_{jk}-\frac12\gamma_{jk}S),
\end{equation}
where $S$ is the trace of $S_{jk}$. First, we can use the junction conditions to extract the background $\Sigma$ and $\Theta$:
\begin{subequations}
    \label{eq:metric jumps}
    \begin{equation}
        [[\sqrt{h}]]=-4\pi\Sigma R,
    \end{equation}
    \begin{equation}
        \Bigr[\Bigr[\frac{f'\sqrt{h}}{f}\Bigr]\Bigr]=8\pi(\Sigma-2\Theta).
    \end{equation}
\end{subequations}
Here, primes denote derivatives with respect to $r$ (and will do so for this entire work) and $R_{\text{shell}}$, which we will frequently abbreviate to $R$, is the areal radial coordinate of the shell. One notable requirement of the junction conditions is that the metric components along the hypersurface which describe the shell must be continuous. This means that $f(r)$ cannot be the canonical Schwarzschild form both inside and outside the shell, as the central masses are different in these regions. To amend this, we take
\begin{equation}
    f(r)=\begin{cases}
        f_{\text{int}}(r)=f_{\text{S}}(M,r),& r<R
        \\
        f_{\text{ext}}(r)=\frac{f_S(M,R)f_{\text{S}}(M+\delta M,r)}{f_S(M+\delta M,R)}, & r>R,
    \end{cases}
\end{equation}
where $\delta M$ is the mass of the shell. The function $h(r)$ can still be of the canonical Schwarzschild form in the interior and exterior regions; the resulting discontinuities in the expressions listed above determine $\Sigma$ and $\Theta$.

From here, we then can compute the discontinuities in the RWG perturbation variables and their derivatives.
Expressions for the perturbations to the stress-energy tensor and extrinsic curvature tensor due to an incident RWG metric perturbation and the resulting junction conditions appear in Appendix A of \cite{PaniGravastar}. Summarizing the main results, the polar sector equations are 
\begin{subequations}
    \label{eq:RW gauge jumps}

    \begin{equation}
        [[K]]=8\pi\sqrt{h}z\Sigma,
    \end{equation}
    \begin{equation}
        [[H]]=8\pi\sqrt{h}z(\Sigma-2\Theta),
    \end{equation}
    \begin{equation}
        \bigjump{\sqrt{h}\Bigr(\frac{H}{R}-K'\Bigr)}+\sqrt{h}z\bigjump{\frac{2h-Rh'}{R^2}}=8\pi\delta\Sigma,
    \end{equation}
    \begin{multline}
        \bigjump{\sqrt{h}\Bigr(K'-H'-\frac{2i\omega H_1}{f}-\frac{H}{R}\Bigr(1+\frac{Rf'}{2f}\Bigr)\Bigr)}
        \\
        +\sqrt{h}z\bigjump{\frac{h'}{R}-\frac{2h}{R^2}+\frac{hf''}{f}-\frac{f'h'}{2f}}=-16\pi \delta\Theta,
    \end{multline}
\end{subequations}
where $\delta\Sigma$ and $\delta\Theta$ are the changes in energy density and surface tension due to the metric perturbation acting on the shell and $z$ is one of a set of gauge variables which are used to convert to a coordinate system in which the shell remains on a worldtube of fixed radius. Note that while $h$ is discontinuous across the shell, the quantity $\sqrt{h}z$ is continuous and thus can be pulled out from the discontinuity terms. 

Eq. \eqref{eq:RW gauge jumps} in principle allows us to solve for $[[Z]]$, $[[Z']]$, $\delta\Sigma$, and $z$, as the RWG perturbation variables $H$ and $K$ and their derivatives can be expressed in terms of $Z$ and $Z'$. We quote standard expressions for the homogeneous case from \cite{HamerlyHorizonDeformations}: 
\begin{widetext}
\begin{subequations}
\begin{equation}
    H_{\text{int}}(r)=-\frac{\lambda^2(\lambda+1)r^3+3\lambda^2Mr^2+9\lambda M^2r+9M^3}{r^2(\lambda r+3M)^2}Z_{\text{int}}(r)
    +\frac{\lambda r^2-\lambda  Mr-3M^2}{r(\lambda r+3M)}\partial_rZ_{\text{int}}(r)
    +rf_{\text{S}}(M,r)\partial^2_rZ_{\text{int}}(r),
    \label{eq: H_int}
\end{equation}
\begin{equation}
    H_{1,\text{int}}(r)=-i\omega \frac{\lambda r^2-3\lambda Mr-3M^2}{rf_{\text{S}}(M,r)(\lambda r+3M)}Z_{\text{int}}(r)-i\omega r\partial_rZ_{\text{int}}(r),
    \label{eq: H1_int}
\end{equation}
\begin{equation}
    K_{\text{int}}(r)=\frac{\lambda(\lambda+1)r^2+3\lambda Mr+6M^2}{r^2(\lambda r+3M)}Z_{\text{int}}(r)+f_{\text{S}}(M,r)\partial_rZ_{\text{int}}(r).
    \label{eq: K_int}
\end{equation}
\label{eq: polar conversions}
\end{subequations}
\end{widetext}
The Zerilli equation can be used to express $\partial^2_rZ$ as a function of $Z$ and $\partial_rZ$. Note that for the waveforms generated in future sections, the metric reconstruction needed for application of the junction conditions is always done in vacuum (i.e., the waveform source function is zero at and beyond the shell radius), and thus the homogeneous expressions are sufficient. The corresponding inhomogeneous forms can still be found in \cite{HamerlyHorizonDeformations}.

However, there are a couple details which must first be accounted for. First, due to the additional prefactor in $f_{\text{ext}}(r)$, the Zerilli equation with time coordinate $t$ and mass $M+\delta M$ will not hold for $Z_{\text{ext}}$ composed from the exterior RWG perturbation variables $H_{\text{ext}}$, $H_{1,\text{ext}}$, and $K_{\text{ext}}$. However, if we define 
\begin{equation}
    \tilde t\equiv t\sqrt{(1-2M/R)/(1-2(M+\delta M)/R)}\equiv \alpha t,
\end{equation}
which gives $f_{\text{ext}}(r)=f_{\text{S}}(M+\delta M,r)\alpha^2$, the exterior metric returns to the Schwarzschild form in the coordinates $(\tilde t,r,\theta,\phi)$. Now the Zerilli equation for $\tilde Z_{\text{ext}}$, composed from $\tilde H_{\text{ext}}$, $\tilde H_{1,\text{ext}}$, and $\tilde K_{\text{ext}}$, should hold with $\partial^2/\partial \tilde t^2$ instead of $\partial^2/\partial t^2$. 

This change in the external time coordinate reduces the wave frequency outside the shell by a factor of $\alpha$. This is akin to the gravitational redshift of waves traveling away from a black hole -- the sudden appearance of additional mass $\delta M$ makes local proper time pass more slowly than if the shell were not present. For the remainder of this work, we use the convention that $\omega$ refers to the frequency of wave solutions inside the shell, and $\tilde\omega\equiv \omega/\alpha$ is the frequency outside the shell (and thus measured by a distant observer).

We finally must connect the new perturbations $\tilde H_{\text{ext}}$, $\tilde H_{1,\text{ext}}$, and $\tilde K_{\text{ext}}$ to the standard perturbations so that we can apply the junction conditions.
To do so, we rely on the fact that the spacetime interval due to the RWG perturbation must be coordinate-invariant. Considering a displacement purely in the time direction, the fact that the conversion from $t$ to $\tilde t$ is a global constant allows us to write $ds^2=g_{\tilde t\tilde t}d\tilde t^2= g_{tt}dt^2$. Expanding this identity produces
\begin{multline}
    f_{\text{S}}(M+\delta M,r)\tilde H_{\text{ext}}(r)Y_{\ell m}(\theta,\phi) d\tilde t^2
    \\
    =f_{\text{ext}}(r)H_{\text{ext}}(r)Y_{\ell m}(\theta,\phi) dt^2,
\end{multline}
which leads to $\tilde H_{\text{ext}}=H_{\text{ext}}$. Since $K_{\text{ext}}$ does not appear in $t$-index components of the metric perturbation, $\tilde K_{\text{ext}}=K_{\text{ext}}$. However,
\begin{equation}
    \tilde H_{1,\text{ext}}(r)Y_{\ell m}(\theta,\phi) d\tilde tdr=H_{1,\text{ext}}(r)Y_{\ell m}(\theta,\phi) dtdr,
\end{equation}
which returns $\tilde H_{1,\text{ext}}=H_{1,\text{ext}}/\alpha$. 
To convert the perturbation variables in the new coordinates $\tilde H_{\text{ext}}(r)$, $\tilde H_{1,\text{ext}}(r)$, and $\tilde K_{\text{ext}}(r)$ into functions of $\tilde Z_{\text{ext}}(r)$, we can use the same expressions written in Eqs. \eqref{eq: H_int}, \eqref{eq: H1_int}, and \eqref{eq: K_int}, after replacing all instances of $M$ by $M+\delta M$ and all instances of $\omega$ by $\tilde\omega$. 

Further clarification on our method for computing the discontinuous changes in the polar master function and its derivative due to the shell appears in Appendix \ref{app: junction conditions}.

In the axial sector, the junction conditions for $h_0(\omega,r)$ and $h_1(\omega,r)$ are simple, and again derived in \cite{PaniGravastar}:
\begin{equation}
    [[h_0]]=[[\sqrt{h}h_1]]=0.
\end{equation}
The conversion between the RWG variables $h_0$ and $h_1$ and the Regge-Wheeler master function $Q$ is simple to derive from Eqs. \eqref{eq: axial perturbation conversions} and \eqref{eq: RW function}. 
Once again, the transformation to the $\tilde t$ time coordinate in the exterior region reduces frequencies by a factor of $\alpha$. Finally, using the invariance of the spacetime interval shows that in this coordinate system, $\tilde h_{0,\text{ext}}=h_{0,\text{ext}}/\alpha$ and $\tilde h_{1,\text{ext}}=h_{1,\text{ext}}$.

The solution to the junction conditions at the shell radius $R$, modified with the changing mass and frequency, is just
\begin{subequations}
    \label{eq: axial jumps solved}
    \begin{equation}
    Q_\text{ext}=Q_\text{int}/\alpha,
    \end{equation}
    \begin{equation}
    Q'_\text{ext}=Q'_\text{int}+\frac{1}{\alpha R}\Bigr(\alpha-1+\frac{2\delta M/R}{f_S(M+\delta M,R)}\Bigr)Q_\text{int}.
    \end{equation}
\end{subequations}
Indeed, as $\delta M\rightarrow0$, $\alpha\rightarrow1$ and thus the axial master function becomes continuous across the boundary.

\subsection{Construction of the Homogeneous Wavefunctions}
\label{subsec: construction}
Having completed our treatment of the effect of the shell on the polar and axial master functions, we now can construct homogeneous wavefunctions on the entire radial domain. Obtaining such functions is essential to computing ringdown properties and waveforms. In this subsection, we will limit our discussion to the polar sector, but the analogous procedure applies cleanly to the axial sector.

In calculating the ringdown behavior of this system, we consider two homogeneous solutions to the Zerilli equation: the ``in" solution, which imposes $Z^{\text{in}}(\omega, r) \to e^{- i \omega r_{*}}$ at the horizon ($r_*\rightarrow-\infty$), and the ``up" solution, requiring $Z^{\text{out}}(\omega, r) \to e^{ i \omega r_{*}}$ as $r_{*} \to \infty$. 
In practice, we must add correction terms to account for starting numerical integration procedures at finite values of $r_*$\cite{Chandrasekhar:1975zza}. Specifically, we define initial conditions for the ``in" (``up") solutions at $r_{*,0}\ll 0$ ($r_{*,0}\gg 0$) via the following expansions\footnote{We arbitrarily choose to truncate these expansions at 10 terms, as additional contributions are very small after this point for our choices of $r_{*,0}$.}:
\begin{subequations}
\begin{equation}
    Z^{\text{in}}(\omega,r_{*,0})=e^{-i\omega r_{*,0}}\Bigr(1+\sum_{n=1}^{10}a_n(r_0-2M)^n\Bigr),
\end{equation}
\begin{equation}
    Z^{\text{up}}(\omega,r_{*,0})=e^{i\omega r_{*,0}}\Bigr(1+\sum_{n=1}^{10}b_nr_0^{-n}\Bigr),
    \label{eq: Z up initial condition correction}
\end{equation}
\end{subequations}
where the coefficients $a_n$ and $b_n$ are chosen to ensure the Zerilli equation is satisfied by the expansion at $r_{*,0}$.

The different masses describing the interior and exterior solutions produce slightly different definitions of the $r_*$ coordinate in the interior and exterior regions. In this work, we generally do not distinguish between these two definitions in our notation, but it can be assumed that when describing the exterior spacetime, the $r_*$ coordinate refers to the definition with mass $M+\delta M$, and when describing the interior spacetime, the $r_*$ coordinate refers to the definition with mass $M$. However, when specifically referring to the radius of the shell, we will denote its location in the interior and exterior tortoise coordinates as $R_*^-$ and $R_*^+$, respectively.

We now briefly summarize the procedure used to construct the homogeneous solutions to the Zerilli equation on the entire radial domain. The steps below specifically outline our method for starting from the near-horizon region and building the ``in" solution outwards, but the analogous method to start from the radiation zone and build the ``up" solution inwards is also valid. 

\begin{enumerate}[i.]
    \item Choose a frequency $\omega$ and fix ingoing initial conditions in the near-horizon region, e.g. $Z^\text{in}_{\text{int}}(r_*)\sim \exp(-i\omega r_*)$ for $r_*\rightarrow-\infty$. Apply the series of corrections to these initial conditions to account for starting the integration at some finite value of $r_*$.
    \item Integrate the solution $Z^\text{in}_{\text{int}}(r_*)$ outwards using the Zerilli equation to the radius of the shell, $R_*^-$, and extract $Z^\text{in}_{\text{int}}(R_*^-)$ and $\partial_{r_*}Z^\text{in}_{\text{int}}(R_*^-)$.
    \item Use the junction conditions and auxiliary expressions outlined in the previous subsection to compute $\tilde Z^\text{in}_{\text{ext}}(R_*^+)$ and $\partial_{r_*}\tilde Z^\text{in}_{\text{ext}}(R_*^+)$.
    \item Using the values obtained in the previous step as the new initial conditions, integrate the exterior solution $\tilde Z^\text{in}_{\text{ext}}(r_*)$ out to some large value of $r_*$ in the radiation zone using the Zerilli equation, accounting for the modified mass $M+\delta M$ and frequency $\omega/\alpha$.
\end{enumerate}

\subsection{Extracting Ringdown Quantities}
Having constructed the ``in" and ``up" solutions on the entire radial domain, we now finally outline how these solutions enable extraction of important ringdown quantities. Once again, we present the relevant information for just the polar sector, but all of the analogous arguments apply for the axial sector. 

The global homogeneous solutions to the Zerilli equation take the following form:
\begin{equation}
    Z^{\text{in}}(\omega,r_*)=\begin{cases}
        e^{-i\omega r_*},&r_*\rightarrow-\infty
        \\
        A_{\text{in}}e^{-i\tilde\omega r_*}+ A_{\text{out}}e^{i\tilde\omega r_*},& r_*\rightarrow +\infty,
    \end{cases}
    \label{eq: global in solution}
\end{equation}
and 
\begin{equation}
    Z^{\text{up}}(\omega,r_*)=\begin{cases}
        B_{\text{in}}e^{-i\omega r_*}
        +B_{\text{out}}e^{i\omega r_*},
        &r_*\rightarrow-\infty
        \\
        e^{i\tilde\omega r_*},& r_*\rightarrow +\infty,
    \end{cases}
    \label{eq: global up solution}
\end{equation}
where we reemphasize that $\tilde\omega\equiv\omega/\alpha$. The coefficients $A_{\text{in}}$, $A_{\text{out}}$, $B_{\text{in}}$, and $B_{\text{out}}$ are functions of the interior solution frequency $\omega$.

We use these solutions to compute the QNM frequencies of the Schwarzschild with shell spacetime. In the $r_*\rightarrow\infty$ limit, the Wronskian between $Z^{\text{in}}$ and $Z^{\text{up}}$ gives
\begin{equation}
    W(Z^{\text{in}},Z^{\text{up}})\equiv Z^{\text{in}}\partial_{r_*}Z^{\text{up}}-Z^{\text{up}}\partial_{r_*}Z^{\text{in}}=2i\tilde\omega A_{\text{in}},
\end{equation}
which can be used to solve for $A_{\text{in}}(\omega)$. We will see that the coefficient $A_{\text{in}}$ itself plays a key role in the ringdown of a Schwarzschild with shell spacetime, but it can also be used to extract the QNM frequencies. 

Since QNM solutions correspond to ingoing boundary conditions at the horizon and outgoing boundary conditions at infinity, the ``in" and ``up" solutions at a QNM frequency will be scalar multiples of one another. Therefore, QNMs are defined as the frequencies for which the Wronskian between the global ``in" and ``up" solutions vanish. In other words, in terms of the interior frequency, the QNMs are identified by the zeroes of $A_{\text{in}}(\omega)$.

To find the QNM frequencies of a given Schwarzschild with shell configuration, we numerically construct the ``in" and ``up" solutions using the procedure described in Sec. \ref{subsec: construction} for a grid of $\omega$ with real part $-0.03/M\leq\omega_R\leq 0.77/M$ and imaginary part $-0.39/M\leq\omega_I\leq 0.01/M$\footnote{In vacuum, QNM frequencies always have $\omega_I<0$, but we extend the grid to positive imaginary parts here to search for instabilities, which would appear with $\omega_I>0$. We did not find any such modes for any of the shell configurations studied in this work.}. We then identify local minima of $|W(Z^{\text{in}},Z^{\text{up}})|$ on the grid and sample frequencies around those local minima with increasing resolution until both a Wronskian with magnitude less than $10^{-3}$ and a frequency resolution of $10^{-8}$ are achieved. Some of these local minima are numerical artifacts; to confirm the validity of a QNM frequency, we evaluate the Wronskian with two different choices of large $r_{*,0}$ for the $Z^{\text{up}}$ solution and retain frequencies for which $|W(Z^{\text{in}},Z^{\text{up}})|$ remains small.

\section{Quasinormal Modes of the Schwarzschild with Shell Spacetime}
\label{sec: qnms}
In this section, we apply the mathematical formalism constructed in Sec. \ref{sec: mathematical construction} to calculate polar and axial QNM frequencies of the Schwarzschild with shell system. We show that the addition of the shell can source up to three types of QNMs, depending on its properties and the choice of parity: a perturbatively migrating fundamental mode, a large set of modes which destabilize the QNM spectrum and physically are the poles of an effective cavity, and a weakly-damped mode associated with the GW-driven ringing of the spherical shell.
Here and throughout the remainder of this work, we commonly consider two classes of shells: one with $\delta M\ll M$ and $R_{\text{shell}}\gg M$, and the other with $R_{\text{shell}}\sim M$ and $\delta M\sim M$.

\subsection{Spectrum of Polar QNMs}
\label{sec: polar qnm frequencies}
\begin{figure}
    \centering
    \includegraphics[width=\linewidth]{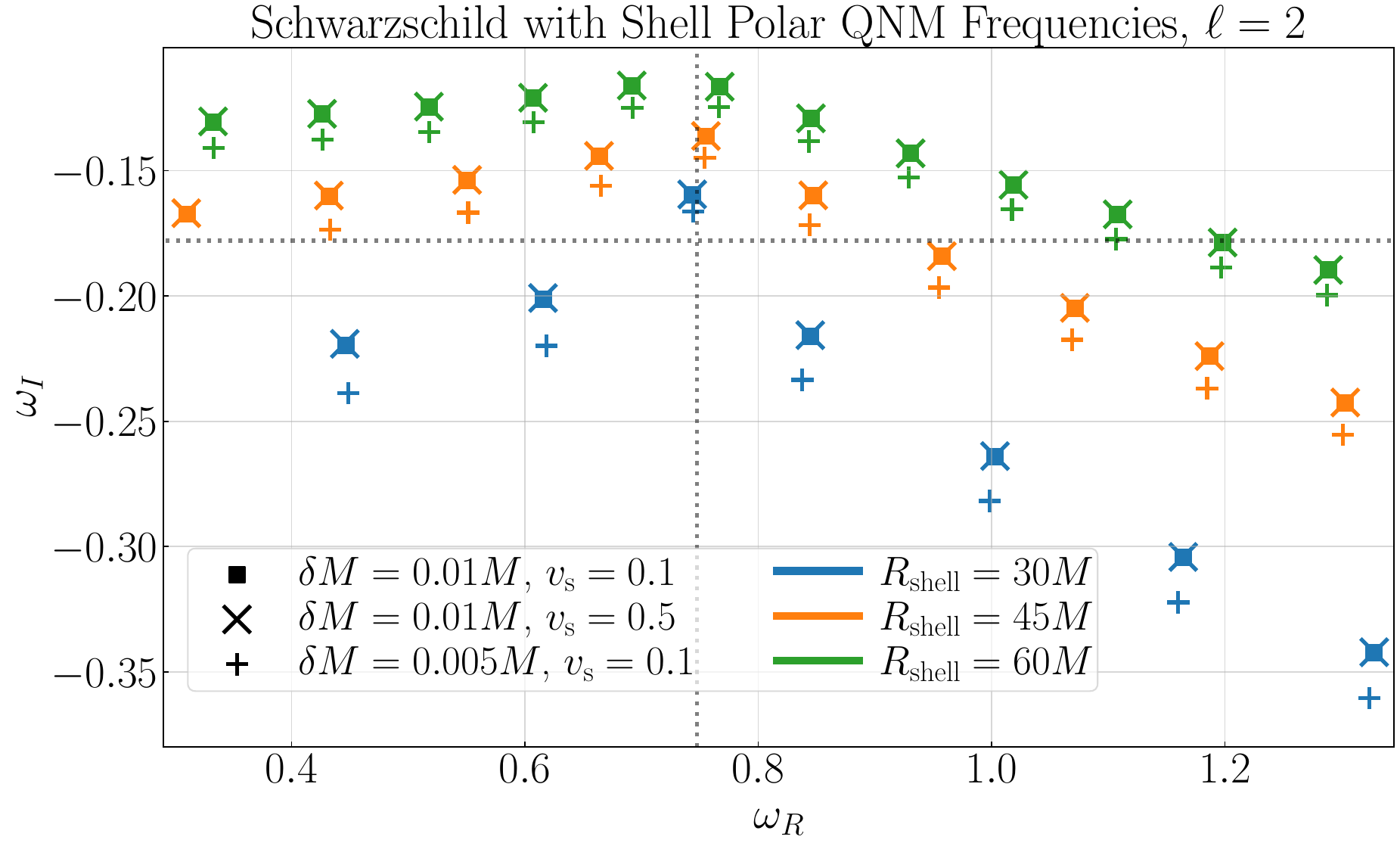}
    \caption{A subset of the $\ell=2$ QNM frequencies for the Schwarzschild with shell spacetime with various choices of $\delta M$, $v_\text{s}$, and $R_{\text{shell}}\gg M$. The vacuum fundamental mode is marked by the intersecting dotted lines. The existence of many modes near the vacuum fundamental mode suggests that the shell destabilizes the QNM spectrum. In the depicted region, the modes are the poles of the effective cavity formed by the curvature potential and shell, with one originating from the migration of the fundamental mode.}
    \label{fig:destabilized spectrum large radii}
\end{figure}

In Fig. \ref{fig:destabilized spectrum large radii}, we present a subset of the QNM frequencies computed for a few choices of systems with $\delta M\ll M$, $R_{\text{shell}}\gg M$. As in many previous works studying the effect of small perturbations placed far from the BH on the system's QNM frequencies, many modes fill the space of complex frequencies where previously there was just one. Furthermore, there is no particular mode which stands out as a displaced analogue of the vacuum fundamental mode. In order to produce this behavior, either \textit{i)} the shell produces new QNMs which have no analogue in the vacuum case, or \textit{ii)} some vacuum modes migrate to an extent far greater than the perturbative parameter, which we will later see can be roughly given by $\delta M/M$. In either case (though the former turns out to be the most apt description), the distribution of QNM frequencies gives adequate evidence to suggest that in the context of \cite{CheungFundamentalInstability}, the fundamental vacuum mode is destabilized by these QNMs, which we henceforce call ``destabilizing" modes. More rigorous definitions of destabilization exist, (e.g., in \cite{JaramilloPseudospectrum}), but we forego such approaches here.

In addition to the frequencies in the destabilized spectrum that appear in Fig. \ref{fig:destabilized spectrum large radii}, the shell sources an additional QNM which exhibits weak damping. This nearly-real ($|\omega_I|\lesssim|\omega_R|/100$) QNM frequency is listed in Table \ref{tab:small mass real qnms} for some of the cases which we studied in Fig. \ref{fig:destabilized spectrum large radii}. We extended our search for frequencies near the real axis up to $\omega_R=10$ but did not find evidence for QNMs beyond the mode listed in Table \ref{tab:small mass real qnms} for each configuration. 

\begin{table}
    \centering
    Nearly-Real Polar QNMs, $\delta M\ll M$, $R_{\text{shell}}\gg M$
    \\
    \vspace*{0.1cm}
    \begin{tabular}{| c |c |}
         \hline
         $R_{\text{shell}}$ & $\omega$, $|\omega_I|\ll |\omega_R|$ ($|\omega_I|\lesssim |\omega_R|/100$)
         \\
         \hline
         $30M$ & $0.0165117-2.38\times 10^{-7}i$
         \\
         \hline
         $45M$ & $0.0106592-9.5\times 10^{-8}i$
         \\
         \hline
         $60M$ & $0.0080346-4.5\times 10^{-8}i$
         \\
         \hline
    \end{tabular}
    \caption{Nearly-real $\ell=2$ QNM frequencies for a Schwarzschild with shell spacetime with shell properties $\delta M=M/100$, $v_{\text{s}}=0.1$, and $R_{\rm shell}\gg M$. Each shell radius results in exactly one such mode.}
    \label{tab:small mass real qnms}
\end{table}
The pattern of QNM frequencies for systems with $\delta M\sim M$, $R_{\text{shell}}\sim M$ differs from that of the previous class of systems. In Table \ref{tab:comparable mass qnms}, we list QNMs for three choices of $R_\text{shell}$ with $\delta M=M/2$. One weakly-damped QNM appears once again, but unlike the previous case, there are fewer modes in proximity to the vacuum fundamental frequency. For $R_\text{shell}=6M$, there is just a single mode which has only slightly migrated away from the vacuum QNM frequency $\omega_0$, and for $R_\text{shell}=8M$, the first of the two listed modes is still clearly associated with a migration of the vacuum fundamental mode. 

However, when the shell approaches a radius of just 10$M$, both listed modes are displaced from $\omega_0$ by similar amounts. A direct association between one of the two listed modes and the vacuum mode cannot be made without information about mode frequencies at other values of $R_\text{shell}$; thus, for this choice of $v_\text{s}$ and $\delta M$, destabilization occurs around $R_\text{shell}\sim 10M$. We provide more details on the behavior of the ``migrating" mode and the destabilizing modes in the following section.  

\begin{table}[t]
    \centering
    Polar QNM Frequencies, $\delta M\sim M$, $R_{\text{shell}}\sim M$
    \\
    \vspace*{0.1cm}
    \begin{tabular}{| c |c | c|}
         \hline
         $R_{\text{shell}}$ & $\omega$, $\omega\approx\omega_0$ & $\omega$, $|\omega_I|\ll |\omega_R|$
         \\
         \hline
         $6M$ & $0.7371959-0.1235047i$ & $0.1287374-2.285\times10^{-5}i$
         \\
         \hline
         $8M$ & $0.6834081-0.1494588i$ & $0.0846855-1.064\times10^{-5}i$ 
         \\
         & $0.8918536-0.2818288i$ &
         \\
         \hline
         $10M$ & $0.6127957-0.1684912i$ & $0.0620124-2.187\times10^{-5}i$
         \\
         & $0.8308987-0.1812928i$ &
         \\
         \hline
    \end{tabular}
    \caption{QNM frequencies of $\ell=2$ modes for a Schwarzschild with shell spacetime with $\delta M=M/2$, $v_{\text{s}}=0.1$,  and $R_{\rm shell}\sim M$. In the convention of \cite{LeaversMethod}, $\omega_0=0.7473433-0.1779247i$. Each configuration features a fundamental mode near $\omega_0$ and exactly one weakly-damped mode, though even as $R_\text{shell}$ approaches just 10$M$, the association of a particular mode with the vacuum fundamental mode starts to become ambiguous, indicating the onset of destabilization.
    }
    \label{tab:comparable mass qnms}
\end{table}
The existence of just one weakly-damped QNM appears to be a general feature for all physically meaningful configurations of the Schwarzschild with shell spacetime\footnote{Interestingly, a study of thin-shell gravastars (which feature a de Sitter interior spacetime) using an identical form of $S_{jk}$ as that of this work (cf. Eq. \eqref{eq: static stress energy}) found that for certain combinations of $v_\text{s}$ and the compactness $\mu$, the weakly-damped QNMs can vanish or even shift to positive $\omega_I$, indicating instabilities \cite{PaniGravastar}. However, these features occur when $v_\text{s}$ and $\mu$ take extreme values; we leave the exploration of such regimes beyond the scope of this paper.}. While we have not presented the results explicitly here, we have confirmed that systems with $\delta M\ll M$, $R_{\text{shell}}\sim M$ and with $\delta M\sim M$, $R_{\text{shell}}\gg M$ also feature exactly one nearly-real QNM frequency.

\subsection{Physical Interpretation of Polar QNM Frequencies}
\label{sec: polar qnms interpretation}
To summarize the previous subsection, we observed three types of QNMs in the Schwarzschild with shell system: a mode which migrates away from vacuum fundamental mode, a set of modes which appear in the space of complex frequencies around the vacuum fundamental mode (the ``destabilizing" modes), and one weakly-damped mode with $|\omega_I|\ll|\omega_R|$. We now offer a physical interpretation for each of these three types of QNM.

\subsubsection{Migrating Mode: $\delta M$ as a Perturbative Variable}
\label{sec: migrating modes}
Despite the destabilization of the QNM spectrum, the fact that our perturbation represents a physical change to the environment suggests that there should be some regime where $\delta M/M$ is small enough such that the new fundamental mode is linearly perturbed away from the vacuum Schwarzschild fundamental QNM frequency $\omega_0$. Indeed, one can show that $A_\text{in}$ receives two perturbations from the existence of the shell -- one from the junction conditions themselves, and one from the reduction in wave frequency by a factor of $\alpha$ -- and both of these perturbations scale with $\delta M$ when $\delta M/M$ is sufficiently small (see Appendix \ref{app: fundamental mode shift}). It then follows that the QNM frequency shift -- that is, the change in frequency from $\omega_0$ needed to cancel out this $\mathcal{O}(\delta M)$ shift and return $A_{\text{in}}$ to zero -- also scales with $\delta M$.   

\begin{figure}
    \centering
    \includegraphics[width=\linewidth]{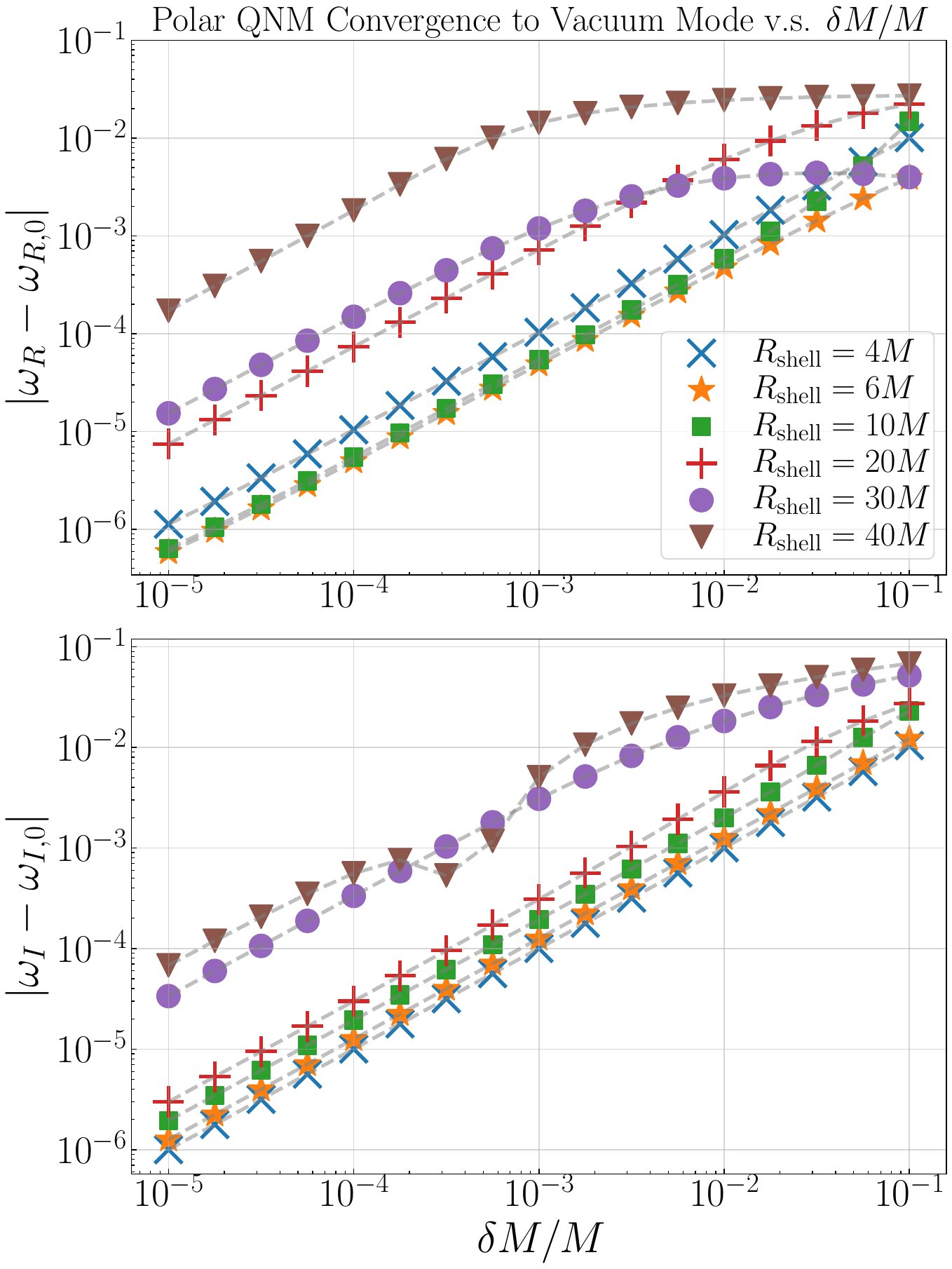}
    \caption{The convergence of the fundamental mode of the Schwarzschild with shell spacetime to the fundamental mode ($\ell=2$) of the vacuum spacetime as the shell mass is reduced towards zero, where $v_\text{s}=0.1$. The top plot depicts $|\omega_R-\omega_{R,0}|$ as a function of $\delta M/M$ for a few choices of $R_{\text{shell}}$, while the bottom plot does the same for the imaginary part, $|\omega_I-\omega_{I,0}|$. For all choices of shell radii, there exists some $\delta M$ such that for shell masses below that value, the QNM frequency is perturbed from $\omega_0$ by an amount proportional to $\delta M$.}
    \label{fig:mode mass trajectory real and imaginary parts}
\end{figure}

We test this claim in Fig. \ref{fig:mode mass trajectory real and imaginary parts}, where we study the convergence of the fundamental mode frequency to $\omega_0$ as a function of $\delta M/M$ and $R_{\text{shell}}$. Regardless of the shell radius, we see that there always exists some perturbative regime where $\omega-\omega_0\propto\delta M/M$. However, as the shell radius grows, the shell mass must become smaller to enter this regime. This behavior occurs due to the exponential growth of the Zerilli wavefunctions for $\omega_I<0$ as the Zerilli equation is integrated radially outwards.

It is worth noting here that we also searched for the effect of the shell on the overtone QNMs -- at least in the perturbative regime, where a mode could be clearly associated with a vacuum overtone in the $\delta M\rightarrow0$ limit -- but were unable to locate any such frequencies. Previous works attempting to locate overtone frequencies via a Wronskian method like the one we employed here have also encountered this difficulty, which likely arises from numerical instabilities occurring when $\omega_I\gtrsim\omega_R$ \cite{Chandrasekhar:1975zza,Molina:2010fb}. 

\subsubsection{Destabilizing Modes: Poles of an Effective Cavity}
The curvature potential and boundary conditions imposed by the shell both can source reflection of incoming waves; therefore, these two features create an effective cavity, within which GWs may be able to resonate. We propose that the QNMs which destabilize the fundamental mode are the poles of this effective cavity. In an analogous form to that proposed in \cite{IanniccariInstability}, they satisfy
\begin{equation}
    1-e^{2i\omega x(\omega)}r_{\text{curvature}}(\omega)r_{\text{shell}}(\omega)=0.
\end{equation}
Here, $x$ is the effective separation in the tortoise coordinate between the reflecting ``surfaces" of the curvature potential and shell (which may depend on $\omega$). The quantities $r_{\text{curvature}}$ and $r_{\text{shell}}$ are the amplitude reflectivities of the curvature potential and the shell to waves incident from the exterior and the interior, respectively. These cavity pole frequencies correspond to the ``trapped modes" identified in \cite{CheungFundamentalInstability}, among previous studies.

The spacing of the QNMs in the destabilized spectra provides a means of testing the interpretation that these modes cycle in the effective cavity. For large shell radii -- that is, when the reflection due to the curvature potential and the shell can be clearly distinguished -- the real part of the spacing between adjacent modes for a given shell configuration is roughly constant (see Fig. \ref{fig:destabilized spectrum large radii}); furthermore, this constant $\Delta\omega_R$ roughly satisfies $\Delta\omega_RR_*\sim \pi$. With this relation, the round-trip of each mode in the cavity accumulates $2\pi$ additional phase relative to the previous mode, maintaining the resonance condition.

The imaginary part of the mode frequencies offers additional confirmation of the cavity interpretation. The segment of the curvature potential outside the shell radius sources a small amount of reflection, even when $\delta M=0$ -- this can be seen in our need for corrections to the Zerilli function at large radii outlined in Eq. \eqref{eq: Z up initial condition correction}. The introduction of the shell sources additional contributions to the reflection coefficient for outward-traveling waves (arising from both the gravitational redshift due to the shell mass and the junction conditions themselves), all of which can be expanded to $\mathcal{O}(\delta M/M)$. 
In other words, the total reflectivity of the exterior portion of the ``cavity" can be written as a constant plus some term $\propto \delta M$.

When the mass of the shell is changed slightly, the QNM frequencies tend to shift by some roughly-constant imaginary value -- e.g., compare the squares and plus signs in Fig. \ref{fig:destabilized spectrum large radii}. With the shell at some large radius, the mode frequencies for shell masses of $\delta M_1$ and $\delta M_2$, which we write here as $\omega_1$ and $\omega_2$, generally satisfy $\delta M_1\exp(-2\text{Im}(\omega_1)R^-_*)\approx\delta M_2\exp(-2\text{Im}(\omega_2)R^-_*)$. Within the cavity interpretation, we understand this feature as the reduction in the wave amplitude reflected by the shell when its mass is reduced from $\delta M_1$ to $\delta M_2$ being made up for by the additional gain in the wave amplitude (due to the imaginary part of the mode frequency) as it propagates through a round-trip inside the cavity. As the mass of the shell is reduced, the destabilizing QNMs therefore continue to move downward in the complex plane. When $\delta M=0$, there is no cavity to resonate in; as expected, these modes disappear -- that is, they indeed have no analogue in the vacuum case. 

\begin{figure}
    \centering
    \includegraphics[width=\linewidth]{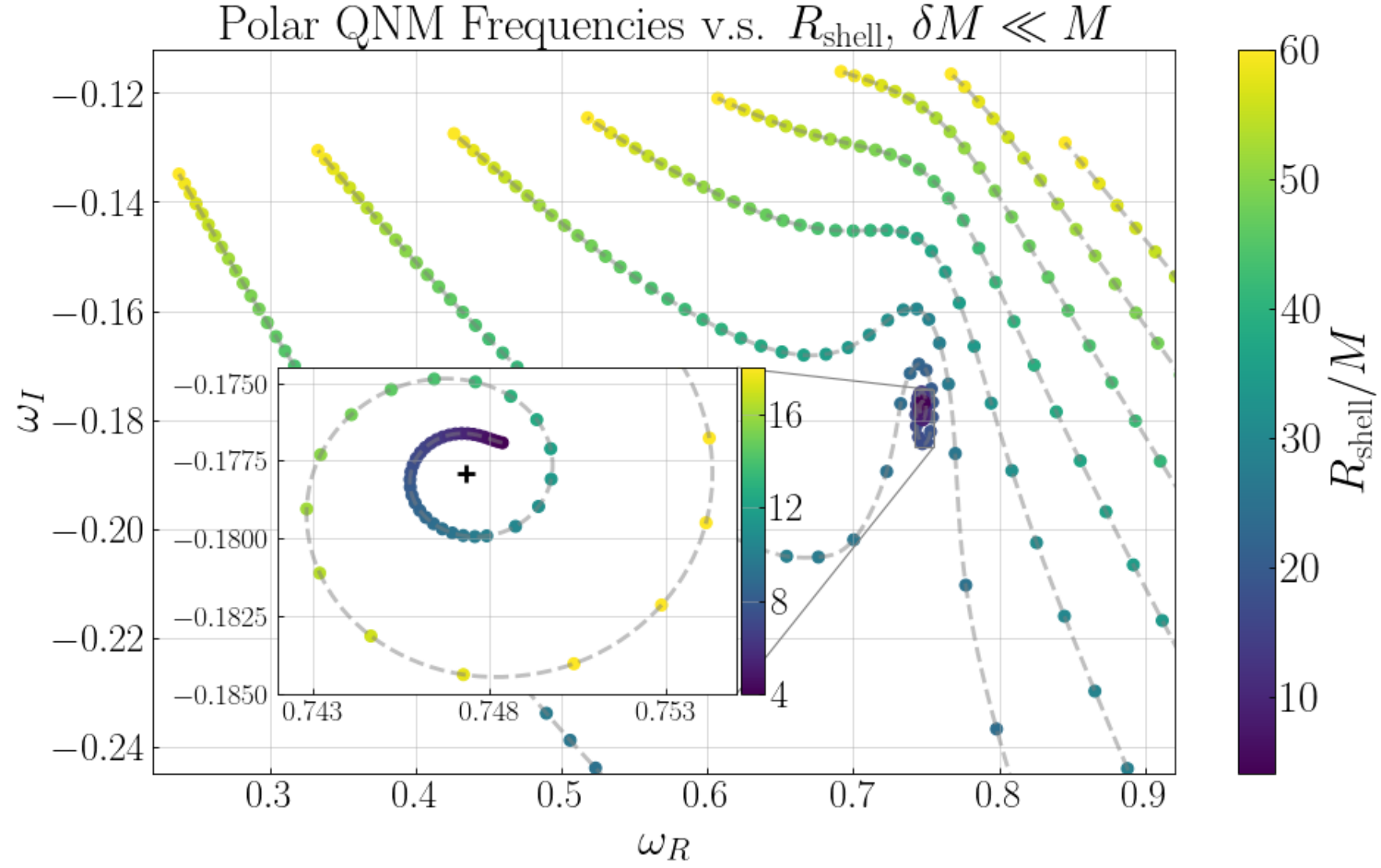}
    \caption{Evolution of a subset of the polar QNM frequencies $\omega_{\text{QNM}}=\omega_R+i\omega_I$ as $R_\text{shell}$ is changed. For this plot, $\ell=2$, $\delta M=M/100$, and $v_{\text{s}}=0.1$. Inset: zooming in on the evolution of the fundamental mode for shell radii less than $18M$. The cross marks the location of the vacuum fundamental QNM frequency $\omega_0$. When $R_\text{shell}\gtrsim 26M$, an ``overtaking instability" occurs: the mode with least negative $\omega_I$ moves to a different track than the one which spirals out from near $\omega_0$.}
    \label{fig:mode radii trajectory full with inset}
\end{figure}

It is also interesting to study how the destabilizing mode frequencies evolve as the radius of the thin shell is varied. In Fig. \ref{fig:mode radii trajectory full with inset}, we follow the frequency of various branches in the destabilized QNM spectrum for shell radii ranging from 4$M$ to 60$M$. The overall structure is reminiscent of previous works which used just a bump in the curvature potential, such as \cite{CheungFundamentalInstability}. The spiral-like feature appearing in the inset when $R_{\text{shell}}$ increases is another general feature of these perturbed systems, which was first noticed in \cite{LeungDirtyBlackHoles}. 

As the shell radius increases, destabilizing modes emerge with large negative $\omega_I$ and move towards the track of the fundamental mode. Within the cavity interpretation, this can be understood as the exponential growth of the wavefunction across the longer cavity requiring a smaller $|\omega_I|$ to make up for the weak shell reflectivity. Eventually, one destabilizing mode moves to a smaller $|\omega_I|$ than that of the track followed by the fundamental mode -- this is the ``overtaking instability" described in \cite{CheungFundamentalInstability}. For the parameters used in Fig. \ref{fig:mode radii trajectory full with inset}, this occurs at $R_{\text{shell}}\approx 26M$. Once this overtaking occurs, the original migrating mode behaves as a destabilizing mode. This can be seen in Fig. \ref{fig:destabilized spectrum large radii}: for $R_{\text{shell}}=45M$ and $60M$, the mode which lies on the track of the original fundamental mode (occurring at $\omega_R\approx 0.43$ and $0.33$, respectively) demonstrates the same spacing patterns observed above for all the other destabilizing modes. To summarize, for a given $\delta M\ll M$, as $R_\text{shell}$ increases, the single migrating mode changes into a destabilizing mode when its value of $\omega_I$ becomes comparable to that of the other destabilizing modes, which themselves only exist in the presence of the effective cavity.

\subsubsection{Weakly-Damped Modes: Resonant Shell Ringing}
\label{sec: weakly damped mode physics}
Finally, we identify the weakly-damped modes as being associated with the secondary GWs sourced when a metric perturbation drives motion of the shell; since the shell lacks a non-gravitational damping mechanism in our model, the ringing persists for a long time. In computing $[[Z]]$ and $[[Z']]$ using the junction conditions, we can also extract the gauge transformation variable $z$. In the context of \cite{PaniGravastar}, this variable roughly describes the radial oscillations of the shell when a metric perturbation proportional to a particular spherical harmonic $Y_{\ell m}$ is incident upon the shell. Calculating $z$ (numerically) as a function of real $\omega$, we find that the shell oscillations are resonantly amplified at frequencies which match the real part of the nearly-real QNM frequencies. The amplified shell motion drives secondary waves which propagate down to the horizon and out to infinity, thus generating a global solution to the Zerilli equation which is both ingoing at the horizon and outgoing at infinity.

By solving the system of equations describing the junction conditions in Sec. \ref{sec: junction conditions}, one can show that in the limit of $\delta M\rightarrow 0^+$, the gauge variable $z$ acquires poles at frequencies $\omega$ solving the following quadratic equation:
\begin{equation}
    P(\omega^2)\equiv P_0+P_2\omega^2+P_4\omega^4=0.
\end{equation}
The coefficients $P_0$, $P_2$, and $P_4$ are functions of just $M$, $\ell$, $R$, and $v_s$. The exact solutions are provided in Appendix \ref{app: z pole frequencies}, and they agree with the results of Table \ref{tab:small mass real qnms} (where $\delta M=M/100$) to less than 1\% error\footnote{The quadratic equation suggests that there exist two real solutions for $\omega^2$. However, only one of them is positive, so just one resonant frequency appears on the positive real axis -- see Appendix \ref{app: z pole frequencies}.}. In short, as $\delta M\rightarrow0$, the real part of the weakly-damped QNM frequencies approaches some nonzero constant -- the resonant frequency of the shell's oscillations.

This resonant frequency exhibits some enlightening scaling relations in two different perturbative regimes. First, in the limit $1\gg M/R\gg v_\text{s}^2$, the resonant frequency reduces to
\begin{equation}
    \omega^2\xrightarrow{1\gg M/R\gg v_\text{s}^2}\frac{M}{2R^3}(\sqrt{\lambda^2+16\lambda+24}-\lambda-4).
\end{equation}
In this limit, the resonant frequency is set by the travel time of gravitational orbits around the shell, indicating that perturbations propagate within the shell on Keplerian trajectories. In the eikonal limit $\ell\gg1$, the resonant frequency becomes independent of $\ell$, with $\omega^2\approx 2M/R^3$. If we instead impose the limit $1\gtrsim v_\text{s}^2\gg M/R$, the resonant frequency becomes
\begin{equation}
    \omega^2\xrightarrow{1\gtrsim v_\text{s}^2\gg M/R}\frac{2v_\text{s}^2(3+\lambda)}{R^2}.
\end{equation}
In this case, the frequency is now set by the travel time of sound waves around the shell -- we infer that in this limit, the shell is governed primarily by acoustic physics rather than by gravity. The eikonal limit here gives $\omega^2\approx v_\text{s}^2\ell^2/R^2$, which is an expected scaling for sound waves traveling around a circular or spherical object.

Meanwhile, as $\delta M\rightarrow 0$, the imaginary part of these mode frequencies becomes proportional to $\delta M$. From a simplified perspective, the motion of the shell is associated with an energy $\propto\delta M$, while in the standard quadrupole description of GWs, energy is radiated away with a luminosity $\propto\delta M^2$. This results in a damping time $\propto \delta M^{-1}$, and thus an imaginary part of the QNM frequency $\propto \delta M$. Thus, the mass of the shell does not substantially impact the frequencies at which its motion is resonantly amplified, but it does play a clear role in the rate at which these oscillations decay.

Previous studies (not limited to \cite{ChandrasekharStarOscillationsIII,KojimaOscillationSpectra,AnderssonOscillationSpectra}) have found similar weakly-damped QNMs in models of relativistic stars. However, this behavior has generally been attributed to the transition in the curvature potential from the interior of the star to the external vacuum region, which creates a local minimum in the potential around which quasibound states can oscillate. It would be interesting to examine whether the boundary conditions at the shell -- namely, the imposition of the junction conditions along with the change in mass from $M$ to $M+\delta M$ and in frequency from $\omega$ to $\omega/\alpha$ -- might source Dirac-delta- or Heaviside-step-function-like features in the potential that would then allow for quasibound states with frequencies matching that of our weakly-damped QNMs.

\subsection{Spectrum of Axial QNMs}
We now consider the behavior of axial-type perturbations in this spacetime. To evolve the homogeneous solutions and compute the axial QNM frequencies, we follow the same general procedure as used to find the polar QNM frequencies. The substantially different form of the junction conditions for polar and axial perturbations (see Sec. \ref{sec: junction conditions}) immediately suggests that the system responds differently to the two types of perturbations and thus admits different QNM frequencies.

\begin{figure}[t]
    \centering
    \hspace*{-0.05\textwidth}
    \includegraphics[width=0.47\textwidth]{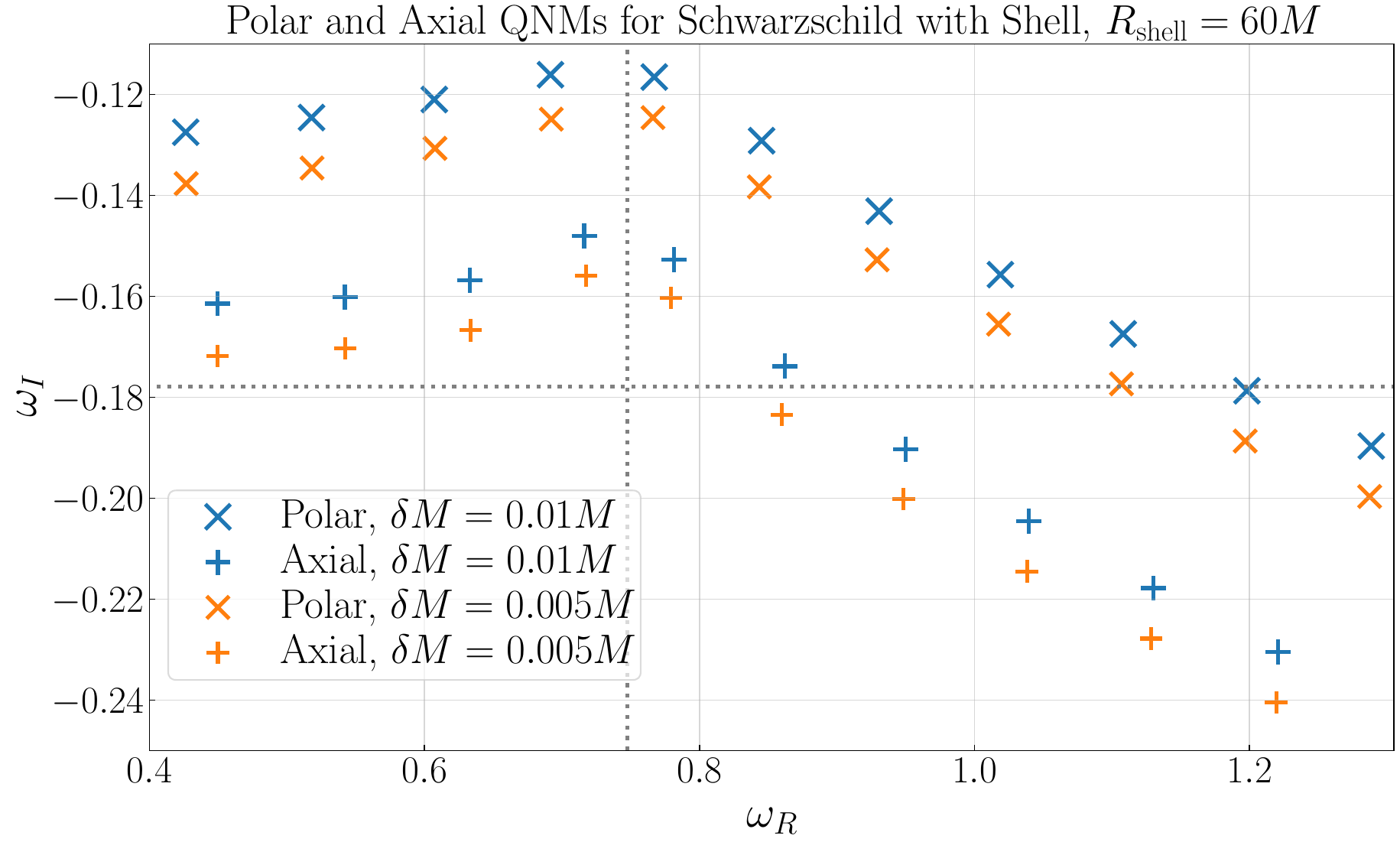}
    \caption{A subset of $\ell=2$ polar and axial QNM frequencies where $\delta M=M/100$, $v_{\text{s}}=0.1$, and $R_{\text{shell}}=60M$. The grey dotted lines mark the location of the vacuum fundamental QNM frequency. The presence of the shell induces isospectrality breaking, though the depicted axial mode frequencies still appear at the poles of an effective cavity.}
    \label{fig: polar vs axial QNMs plot}
\end{figure}

In Fig. \ref{fig: polar vs axial QNMs plot} and Table \ref{tab: polar vs axial QNMs table}, we present polar and axial QNM frequencies for example cases with $\delta M\sim M$, $R_{\text{shell}}\sim M$ and $\delta M\ll M$, $R_{\text{shell}}\gg M$. It is immediately apparent that isospectrality is broken, in that the polar and axial sectors do not feature the same QNM frequencies, but a few more specific features of these results are especially notable.

\begin{table}[t]
    \centering
    \begin{tabular}{|c|c|c|}
        \hline 
        \multicolumn{3}{|c|}{$\delta M\sim M$, $R_{\text{shell}}\sim M$}
        \\
        \hline
        Parity & $\omega$, $\omega\approx\omega_0$ & $\omega$, $|\omega_I|\lesssim |\omega_R|/100$ \\
        \hline
        Polar & $0.7371959-0.1235047i$ & $0.1287374-2.285\times10^{-5}i$ \\
        \hline
        Axial & $0.7655193-0.1742251i$ & None \\
        \hline
        \multicolumn{3}{|c|}{$\delta M\ll M$, $R_\text{shell}\gg M$} \\
        \hline
        Parity & $\omega$, $\omega\approx\omega_0$ &  $\omega$, $|\omega_I|\lesssim |\omega_R|/100$ \\
            \hline
        Polar & None & $0.0080346-4.5\times 10^{-8}i$ \\ 
        \hline
        Axial & None & None \\
        \hline 
    \end{tabular}
    \caption{Polar and axial QNM frequencies which lie near the vacuum frequency $\omega_0$ or nearly on the real axis for shells with properties $\delta M=M/2$, $v_\text{s}=0.1$, $R_{\text{shell}}=6M$ (top rows) and $\delta M=M/100$, $v_\text{s}=0.1$, $R_\text{shell}=60M$  (bottom rows). The shell fails to create axial QNMs with $|\omega_I|\ll|\omega_R|$, and furthermore, when the shell is far from the horizon, the spectrum is destabilized (see Fig. \ref{fig: polar vs axial QNMs plot}) and thus no modes appear near $\omega_0$. All polar results are reproduced from Sec. \ref{sec: polar qnm frequencies}.}
    \label{tab: polar vs axial QNMs table}    
\end{table}

Perhaps the most striking feature is that the axial sector lacks the weakly-damped QNMs observed in the polar sector. After imposing the fixed-worldtube gauge for the shell, the junction conditions account for the GW-driven motion of infinitesimal elements of the shell of matter. The axial junction conditions described in Sec. \ref{sec: junction conditions} are independent of $v_{\text{s}}$, suggesting that the axial perturbations do not change the local matter density in the shell -- that is, they do not create additional stress in the shell. As such, the axial (odd) waves do not perturb the spherically-symmetric (even) matter distribution of the shell in a way that sources secondary GWs. Indeed, the lack of coupling between axial waves and spherically-symmetric matter distributions has been long understood in the context of perturbations to relativistic stars \cite{ThorneNonradialPulsation,ChandrasekharStarOscillationsIV}. Therefore, there are no real-frequency resonances in the shell's response to axial perturbations which could send amplified waves to both the horizon and infinity and thus appear as a QNM.

Despite the lack of coupling between the axial perturbations and the shell matter, the mass of the shell can still act as a perturbative parameter in the migration of the fundamental mode. In fact, the arguments presented in Appendix \ref{app: fundamental mode shift}, which show that in certain limits, the polar fundamental mode frequency shift should scale with $\delta M/M$, can be applied to the axial sector as well. We see evidence of this in Table \ref{tab: polar vs axial QNMs table}, where the axial sector features a QNM near the vacuum $\omega_0$ for the same shell configuration which generated a migrating polar mode. 

Furthermore, the shell can also source a non-zero reflectivity for incident axial metric perturbations. From the solutions to the axial junction conditions in Eq. \ref{eq: axial jumps solved}, we can see that for any $\delta M>0$, the Regge-Wheeler function and its derivative are discontinuous across the boundary, necessarily generating some non-zero reflected amplitude. Due to this reflection, the destabilizing modes observed in the polar sector should appear in the axial sector as well, and their frequencies should lie at the poles of the analogous effective cavity. Indeed, the spacing in between QNM frequencies in the axial sector for the case $R_{\text{shell}}\gg M$, both in between adjacent modes for a single shell configuration and for any given mode between two different shell masses, is remarkably similar to that of the polar sector. The fact that the axial destabilizing modes have substantially more negative imaginary components than their polar counterparts can be understood within the cavity interpretation as arising from the shell reflecting axial waves more weakly than polar waves, a feature itself emerging due to the lack of coupling between axial perturbations and the shell matter.

Notably, the presence of destabilizing modes (and thus a destabilized QNM spectrum) in the axial sector for the shell system contrasts with treatments of axial mode propagation in the presence of extended external matter profiles. These works have rather found a stable QNM spectrum where the frequency shifts are well-characterized by a pure gravitational redshift \cite{PezzellaHaloQNMs}. 

Interestingly, the discontinuity in axial propagation at the shell, which sources the reflectivity that in turn generates the observed destabilizing modes, might in principle also produce features in the potential that would allow for the weakly-damped quasibound states described in \cite{AnderssonOscillationSpectra,KojimaOscillationSpectra,ChandrasekharStarOscillationsIII}, identified for relativistic stars in both the polar and axial sectors. The lack of weakly-damped axial QNMs in our model suggests that perhaps the quasibound state picture does not tell the entire story regarding weakly-damped QNMs of a Schwarzschild with shell spacetime.

\subsection{Summary of QNM Features}
In the preceding subsections, we thoroughly analyzed the QNMs of a Schwarzschild with shell spacetime. For reference, we briefly summarize our results thus far. 

\begin{enumerate}[i.]
    \item For all shell radii, in both the polar and axial sectors, there exists some $\delta M/M$ below which the fundamental mode migrates away from the vacuum frequency by an amount proportional to $\delta M$ -- that is, there always exists some perturbative regime in $\delta M$.
    \item In both the polar and axial sectors, as $R_{\text{shell}}\gg M$ increases for a given $\delta M$, the quasinormal spectrum becomes dominated by the destabilizing modes, which lie at the poles of the effective cavity formed by the curvature potential and the shell.
    \item In the polar sector alone, the coupling between the metric perturbations and the matter of the shell produces weakly-damped QNMs ($|\omega_I|\ll|\omega_R|$). These QNMs appear at the resonant frequencies of the shell's motion, and thus correspond physically to the secondary GWs produced by the shell's amplified ringing.
\end{enumerate}

\section{Ringdown Waveforms and QNM Fits}
\label{sec: ringdown fits}
The primary tool for estimating the properties of a remnant black hole from the ringdown portion of a compact object merger GW signal is extracting the QNM frequencies from the ringdown and mapping them to the remnant mass and spin \cite{IsiRingdownPE,WangRingdownPE,BertiRingdownPE,SiegelRingdownPE}. In order to determine whether this procedure might be generally applicable to parameter estimation for Schwarzschild with shell systems, we now wish to examine how the QNM frequencies computed in Sec. \ref{sec: qnms} are imprinted on true ringdown waveforms. In particular, we are interested in understanding the shell signatures that render the waveform distinguishable from vacuum ringdown signals. 

To carry out this comparison, we compute the waveform $\mathcal{Z}/m$ measured by a distant observer, well beyond the shell radius, generated by a particle of mass $m$ moving on a radial infall trajectory into a BH surrounded by a shell. The radial infall trajectory generates an entirely polar source term -- we choose this to emphasize the influence of the shell, as we have seen that polar sector perturbations couple to the matter in the shell whereas axial sector perturbations do not. In all of our waveforms, the particle begins at rest at a radius of $5M$, which lies inside all shell radii considered. This ensures that the shell only affects wave propagation and not the dynamics of the infall trajectory. Our procedure for generating the resulting waveform is described in Appendix~\ref{app: radial infall}.

\begin{figure}
    \centering
    \includegraphics[width=\linewidth]{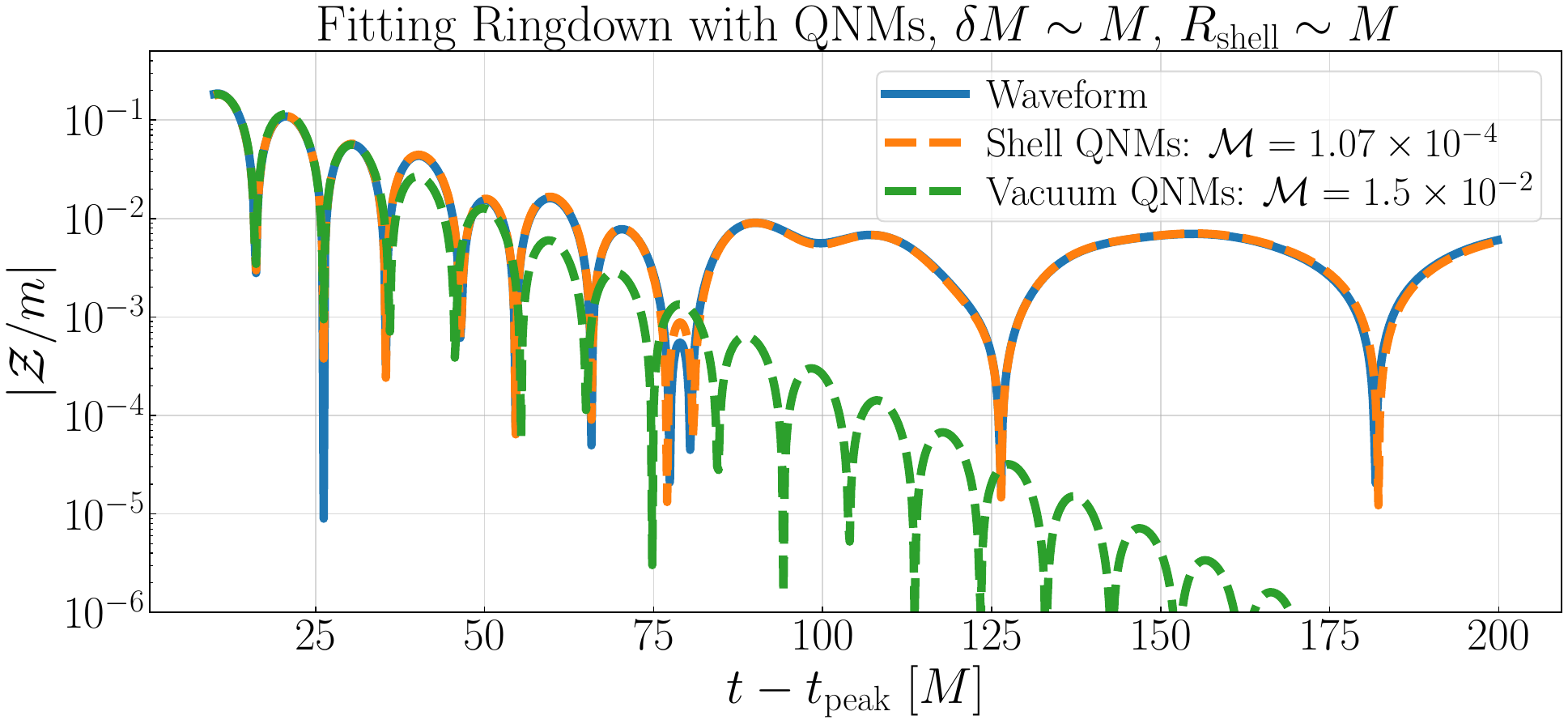}
    \\
    \vspace*{0.1cm}
    \includegraphics[width=\linewidth]{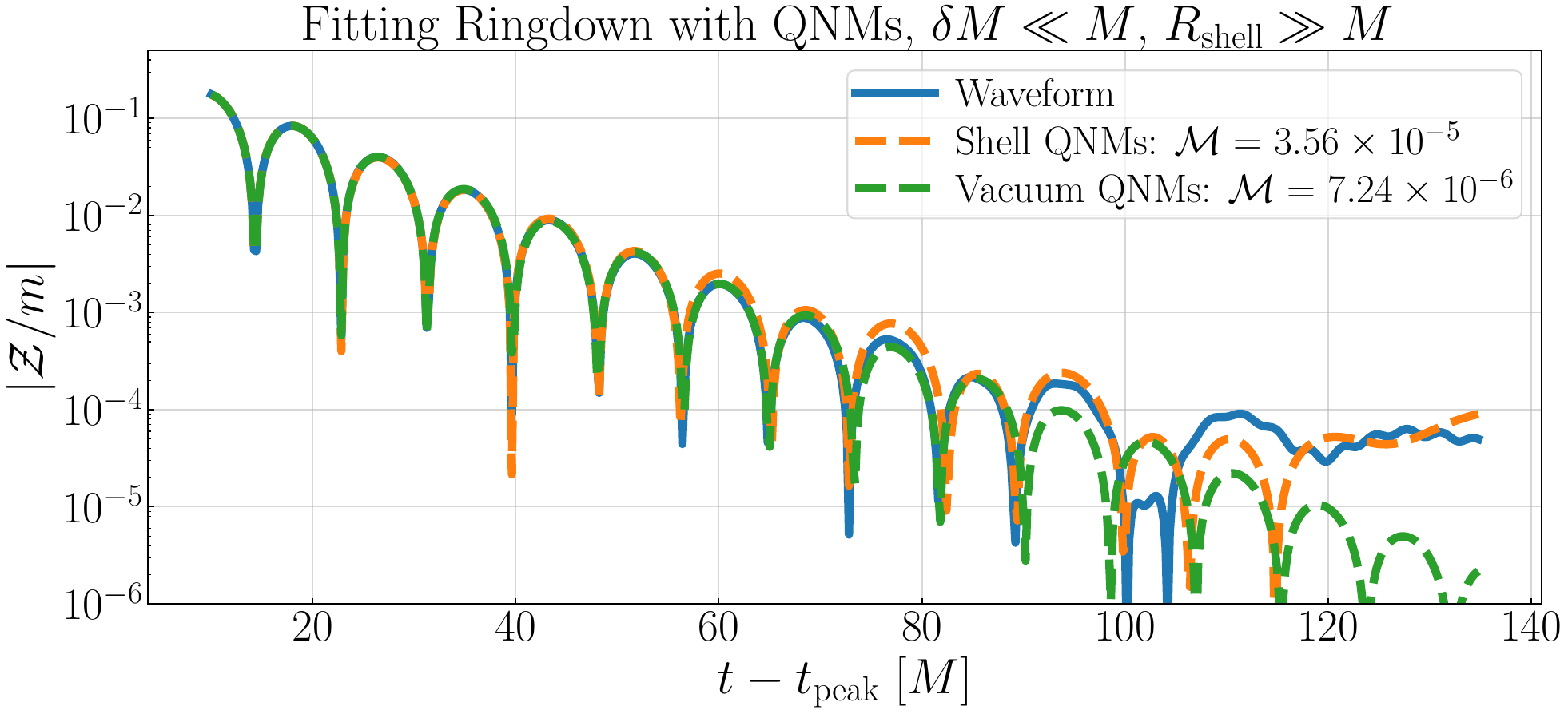}
    \caption{Radial-infall waveforms in the Schwarzschild with shell spacetime along with least square fits using the shell QNM frequencies obtained in Sec. \ref{sec: polar qnm frequencies} as well as the Schwarzschild vacuum QNM frequencies for a BH with mass $M$. We also list the time-domain mismatch, defined in Eq. \eqref{eq: time domain mismatch}, for each QNM fit. Top: $\delta M=M/2$, $v_{\text{s}}=0.1$, $R_{\text{shell}}=6M$. Bottom: $\delta M=M/100$, $v_{\text{s}}=0.1$, $R_{\text{shell}}=30M$. In the former case, the shell ringing sets in shortly after the start of ringdown, and the shell QNMs fit the ringdown far more cleanly than the vacuum QNMs. In the latter case, the shell ringing does not appear, and the early ringdown is well described by the vacuum modes. Notably, even just a few modes of the destabilized spectrum can still fit the early ringdown to decent precision.}
    \label{fig:radial infall ringdown qnm fits}
\end{figure}

With the waveforms in hand, we wish to determine whether the shell system QNM frequencies provide a substantially better fit for the ringdown waveforms than the vacuum QNMs, which would suggest that the program of parameter estimation through QNM extraction could be extended to this class of systems. In Fig. \ref{fig:radial infall ringdown qnm fits}, we present the radial infall waveform with both $\delta M\sim M$, $R_{\text{shell}}\sim M$ and $\delta M\ll M$, $R_{\text{shell}}\gg M$ along with least-squares amplitude fits with two different sets of QNMs: the frequencies of the shell system found in Sec. \ref{sec: polar qnm frequencies} and the vacuum frequencies for a BH of mass $M$, which have been computed by Leaver \cite{LeaversMethod}. We begin the fits at a retarded time of $t=10M$ after the peak of $|\mathcal{Z}|$ and generally end at $t=100M$ after the peak\footnote{In the case of $\delta M\sim M$, $R_\text{shell}\sim M$, we extend the fit to $200M$ after peak strain, as the strong shell ringing well exceeds the level of numerical noise.}. To ensure a reasonable comparison, we use the same number of frequencies (and thus the same number of degrees of freedom) for both the vacuum and shell fits. We choose the number of frequencies as $\text{min}\{N,4\}$, where $N$ is the number of modes found for the given shell system (in the $\ell=2$ harmonic), and convert all QNM frequencies to observed frequencies by dividing by $\alpha$.

For each set of frequencies, we compute the time-domain mismatch $\mathcal{M}$ between the sourced waveform and the sum of QNMs with their amplitudes generated by the least-squares fit. The mismatch is given by
\begin{equation}
    1-\mathcal{M}\equiv \frac{N(\mathcal{Z},\mathcal{Z}_{\text{fit}})}{\sqrt{N(\mathcal{Z},\mathcal{Z})N(\mathcal{Z}_{\text{fit}},\mathcal{Z}_{\text{fit}})}},
    \label{eq: time domain mismatch}
\end{equation}
where $\mathcal{Z}$ is the ringdown waveform, $\mathcal{Z}_{\text{fit}}$ is the sum of QNMs with the least-squares fit parameters, and
\begin{equation}
    N(\mathcal{Z}_a,\mathcal{Z}_b)\equiv\int_{t_{\text{peak}}+10M}^{t_{\text{peak}}+100M}\mathcal{Z}_a(t)\mathcal{Z}_b(t)dt,
\end{equation}
with $t_{\text{peak}}$ being the time of peak strain. We start the integration 10$M$ after peak to allow effects from the prompt response to become subdominant.

In the case where $\delta M\sim M$, $R_{\text{shell}}\sim M$, the ringdown waveform clearly displays a weakly-damped ringing at late times, whose frequency and damping time of this ringing signal are consistent with that of the system's weakly-damped QNM. A similar slowly-decaying component, also associated with the coupling between the polar perturbations and external matter, has been identified for the case of an extended mass distribution outside the BH \cite{CardosoHaloPerturbationFormulas}, but this component generally has not been associated with QNMs of the modified spacetime. Since the shell ringing is strong (only $\sim$ 1 1/2 orders of magnitude weaker than peak strain), the nearly-real polar QNMs can fit the late-time shell ringing in the waveform far more cleanly than the vacuum frequencies, resulting in a significantly smaller mismatch for the shell QNMs. 

We have observed that the shell ringing amplitude scales with $\delta M$ for a given $R_\text{shell}$ and $v_\text{s}$. This behavior can be understood in the context of the structure of $A_\text{in,shell}$ on the real axis, which we explore in Sec. \ref{sec: Ain signatures in ringdown}. It is also consistent with the standard quadrupole picture we outlined in our physical interpretation of the weakly-damped QNM frequencies -- the strain scales as $h\propto \delta M$ while the luminosity scales as $L_\text{GW}\sim \dot h^2 \propto \delta M^2$.

\begin{figure}
    \centering
    \includegraphics[width=\linewidth]{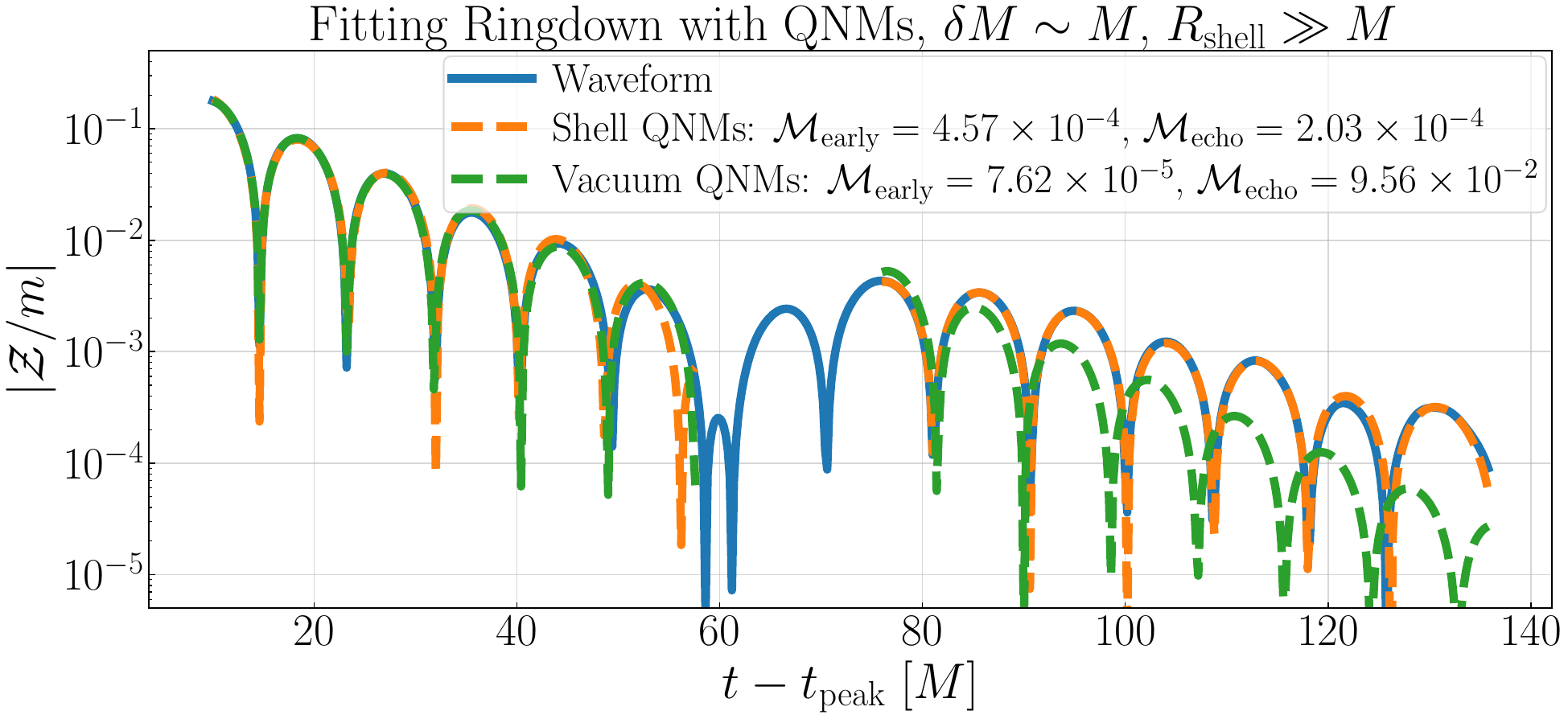}
    \caption{Identical to Fig. \ref{fig:radial infall ringdown qnm fits}, except we have $\delta M=M/2$, $v_{\text{s}}=0.1$, $R_{\text{shell}}=30M$. We also have performed least-squares fits on both the early ringdown ($\sim[10M,55M]$ after $t_\text{peak}$) and the echo ($\sim [75M,135M]$ after $t_\text{peak}$). The time-domain mismatch on the early ringdown $\mathcal{M}_\text{early}$ indicates that the vacuum QNMs still outperform the shell QNMs at early times, but the QNM fits on the echo produce a substantially better mismatch $\mathcal{M}_\text{echo}$ when using the shell QNM frequencies.}
    \label{fig:radial infall echo}
\end{figure}

In the case where $\delta M\ll M$, $R_{\text{shell}}\gg M$, the shell ringing is much weaker, with no clear component appearing above the numerical noise floor appearing around $10^{-4}$. Of course, the smaller mass of the shell reduces the magnitude of the ringing, but even when $\delta M\sim M$ for this choice of $R_\text{shell}$ (plotted in Fig. \ref{fig:radial infall echo}, which we will introduce shortly), there is no weakly-damped component visible at late times. The shell ringing amplitude is attenuated far more rapidly by increasing $R_\text{shell}$ than by decreasing $\delta M$, primarily because $R_\text{shell}$ sets the resonant frequency -- we provide further explanation of this behavior in Sec. \ref{sec: Ain signatures in ringdown}.

After carrying out the QNM fits on this waveform, we see that the vacuum QNMs produce a mismatch which outperforms that of the shell QNMs by a factor of $\sim 6$. From this result, we conclude that in the presence of the shell, the ringdown at sufficiently early times is still well described by vacuum QNMs. This suggests that the early time domain ringdown is generally stable against the addition of a shell -- as long as the shell is sufficiently light and/or far away -- in line with the results of studies of bumps in the curvature potential. Furthermore, we observe that even though the QNM spectrum is destabilized when $R_\text{shell}\gg M$, and thus features no modes which lie very near the dominant vacuum fundamental mode, even a small number of the shell system QNMs can still reproduce the ringdown signature of the vacuum system.

We also considered the least-squares fits using the set of vacuum QNMs for a Schwarzschild BH of mass $M+\delta M$. These QNMs produced mismatches similar to that of the vacuum QNMs for a BH with a mass $M$, so for the sake of clarity, we have omitted those results in Fig. \ref{fig:radial infall ringdown qnm fits}.

Previous works which have found stability in time-domain ringdown signals against perturbations directly applied to the curvature potential (e.g. \cite{BertiTimeDomainStability}) have suggested that in light of the destabilized QNM spectrum, a discrete set of frequencies may not be the most appropriate means for parameterizing ringdown signatures in general merger environments. Our complete physical treatment of polar modes reveals behavior which did not appear in those studies. But despite the shell QNMs faithfully matching the ringdown in the $\delta M\sim M$, $R_{\text{shell}}\sim M$ case, the fact that the destabilized spectrum can masquerade as a vacuum ringdown at early times in the $\delta M\ll M$, $R_{\text{shell}}\gg M$ case lends additional credence to the claim of \cite{BertiTimeDomainStability} regarding the need for caution in parameterizing ringdown signatures solely through QNMs.

Many previous studies of ringdown waveforms with bumps in the curvature have observed echoes at late times due to reflection off the bump feature. The cavity interpretation of the destabilizing mode frequencies immediately suggests that generating such echoes should be possible in the shell system as well. We did not observe echoes in the waveform with $\delta M\sim M$, $R_\text{shell}\sim M$ because the shell ringing is too strong, overpowering any echo which could appear at late times, and we did not observe echoes when $\delta M\ll M$, $R_\text{shell}\gg M$ since the shell reflectivity (which we have seen is dictated by $\delta M$) is too small to generate an echo strong enough to see above the numerical noise.

If we use a system with $\delta M\sim M$, $R_\text{shell}\gg M$, however, the reflectivity of the shell could be large enough to produce an echo, while the shell resonant frequency could be small enough for the shell ringing mode to only be weakly excited in the waveform (see Sec. \ref{sec: Ain signatures in ringdown}). In Fig. \ref{fig:radial infall echo}, we produce the radial infall waveform for a shell with such properties (namely, $\delta M=M/2$, $R_\text{shell}=30M$). At a time of around $75M$ after peak strain, a second peak in $|\mathcal{Z}|$ occurs, followed by a ringdown signature -- clear evidence of an echo. Note that while the echo arrives slightly later than the naively-expected time of $2R_*\approx 70M$ after peak strain, we have confirmed that the difference in echo arrival time for two different large shell radii is twice the difference in their tortoise radial coordinates, squarely in line with the picture of echoes emerging from reflections off the shell and curvature potential.

We also performed two least-squares QNM fits on this waveform: one on the early ringdown ($\sim[10M,55M]$ after peak strain) and one on the echo ($\sim [75M,135M]$ after peak strain). The region $\sim[55M, 75M]$ corresponds roughly to the ring-up of the echo. The time domain mismatch on the early ringdown $\mathcal{M}_\text{early}$ shows that this window is still well described by the vacuum QNMs, matching our previous results. If we perform a fit on the complete range $\sim[10M,135M]$ after peak strain, the vacuum modes generate a lower mismatch due to their superiority during the early ringdown, when the strain is larger, but the mismatch is smaller that of the shell QNM frequencies by only a factor of $\sim 2$ instead of the factor of $6$ seen in the fits on the $\delta M\ll M$, $R_\text{shell}\gg M$ waveform.

Remarkably, though, the shell QNMs provide a much better fit to the echo component, generating a mismatch $\mathcal{M}_\text{echo}$ which is almost three orders of magnitude better than that of the vacuum QNMs. One possible explanation for this behavior is that the echo is a consequence of the same physics (namely, the reflectivity of the shell) which generates the destabilizing modes, and thus its morphology should be well described by such QNMs. Meanwhile, the early ringdown is primarily driven by the existence of the extended curvature potential which generates the vacuum modes, and thus that part of the waveform is well described by the vacuum QNMs. A more concrete formulation of this hypothesis, along with further exploration of the role of the shell and vacuum QNMs at different times in these waveforms, is certainly warranted in future work.

Beyond the waveforms with pure polar source terms, it also could be interesting to examine whether waveforms with non-zero source terms in the axial sector, which notably does not feature weakly-damped modes, might therefore carry signatures of echoes from the shell regardless of the shell parameters. 

It is apparent that across the range of shell radii, substantial deviation to the early polar ringdown waveform is only achieved when the shell mass is comparable to or larger than the mass of the black hole. Therefore, we conclude that much like bumps in the curvature potential, early ringdown in the polar sector is stable under the addition of a shell with $\delta M\ll M$ to the local environment. However, if the ringdown at sufficiently late times can be observed, features in the waveform which clearly point to a deviation from the purely vacuum spacetime may be identified from QNM fits.

\section{Structure of $A_{\text{in}}(\omega)$}
\label{sec: Ain on real axis}
Because QNM frequencies appear as poles of $1/A_{\text{in}}(\omega)$ in the complex frequency plane, the fact that sourced time-domain ringdown waveforms appear as a sum of QNMs implies that the structure of these poles must influence the behavior of $1/A_{\text{in}}(\omega)$ (a component of the Green's function for sourced ringdown, see Appendix \ref{app: radial infall}) on the real frequency axis. Therefore, if the addition of some bump to the curvature potential perturbs $1/A_{\text{in}}(\omega)$ in such a way that the locations of its poles are drastically shifted while its structure is changed stably on the real axis, then the ringdown waveform should maintain its correspondence with the QNM frequencies of the unperturbed potential. Indeed, when the QNM frequencies are destabilized by some small bump in the Regge-Wheeler or Zerilli potentials placed far from the central BH, the greybody factor (defined by $\Gamma(\omega)\equiv |A_{\text{in}}(\omega)|^{-2}$) on the real axis is altered in a manner such that the early ringdown waveform displays stability to the perturbation \cite{OshitaGreybodyI,OshitaGreybodyII,OshitaGreybodyStability,PaniGreybodyStability,BertiTimeDomainStability}.

Given that the early ringdown in the polar sector appears stable under the addition of the shell, we wish to determine whether the stability can be inferred from the structure of $A_{\text{in}}(\omega)$ on the real axis. We first consider how the shell of matter affects this quantity in the polar sector. For the remainder of this work, we use $A_\text{in}$ and $A_\text{in,shell}$ interchangeably to describe the ingoing wave amplitude at infinity for the ``in" solution in the shell system, and use $A_\text{in,vacuum}$ exclusively to describe this quantity when the shell is absent.

\subsection{$A_{\text{in}}$ in the Polar Sector}
\label{sec: Ain polar sector}
\begin{figure}
    \centering
    \includegraphics[width=\linewidth]{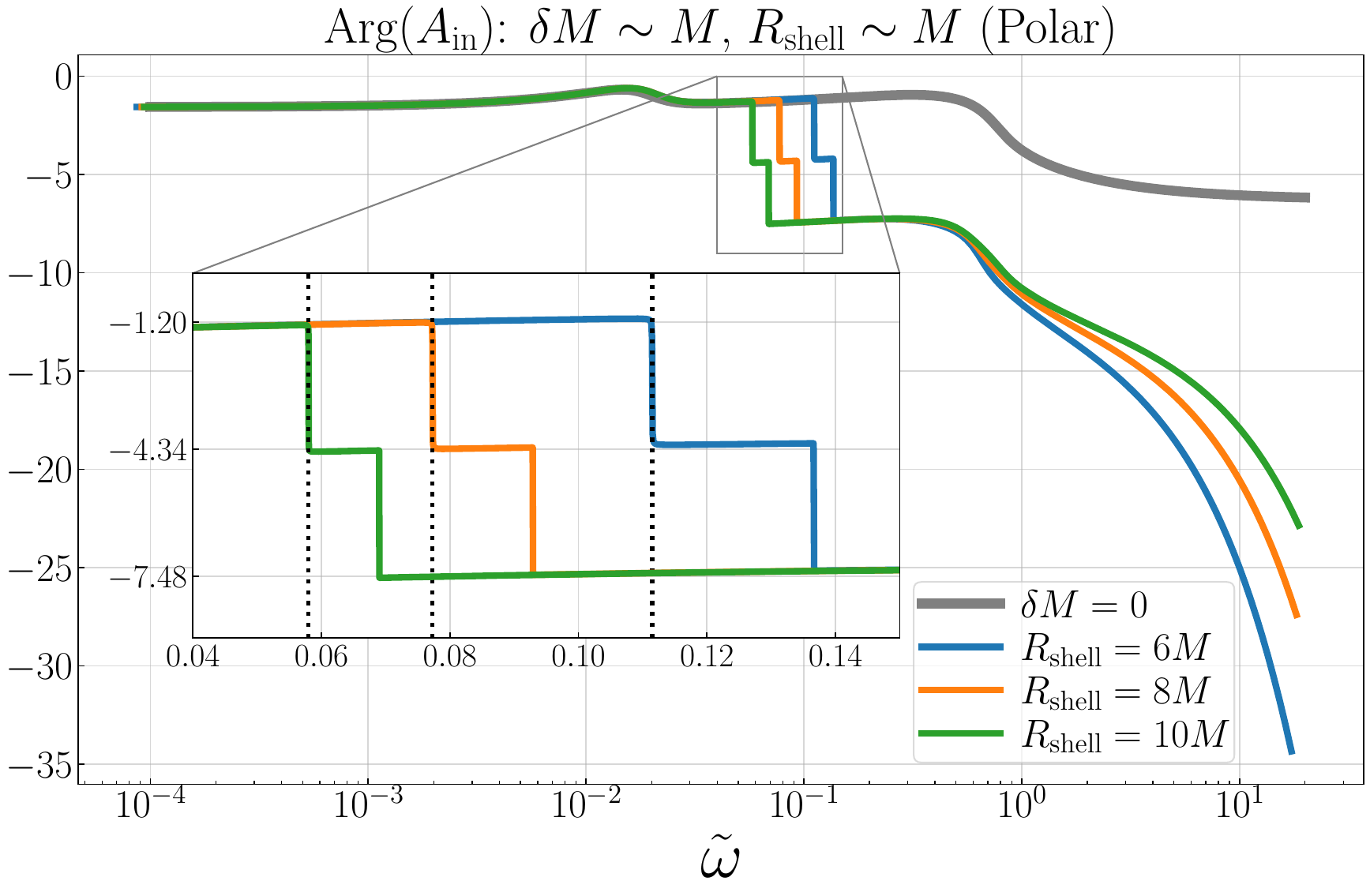}
    \includegraphics[width=\linewidth]{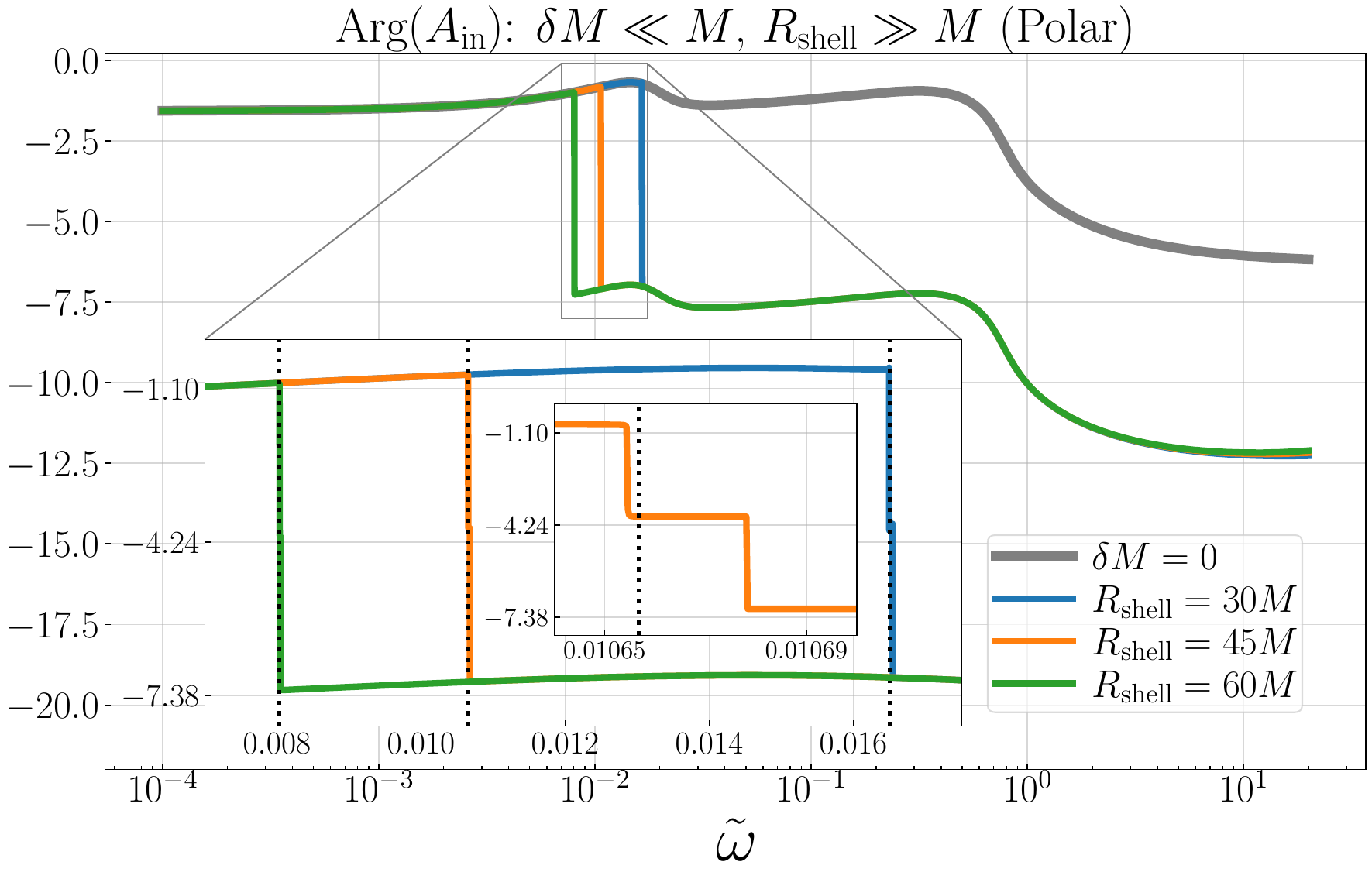}
    \caption{Top: the $\ell=2$ polar $\phi(\tilde\omega)\equiv
    \text{Arg}(A_{\text{in}}(\tilde\omega))$ for a Schwarzschild with shell system with $\delta M=M/2$, $v_{\text{s}}=0.1$, and three choices of $R_{\text{shell}}\sim M$. Bottom: the same quantity, but with $\delta M=M/100$ and three choices of $R_{\text{shell}}\gg M$. Dotted black lines are placed at the real parts of the nearly-real QNM frequencies identified in Sec. \ref{sec: polar qnm frequencies} (divided by $\alpha$ to convert to observed frequency $\tilde\omega$). The behavior of $\phi(\tilde\omega)$ generally matches that of the vacuum system (gray), but at the real part of the weakly-damped QNMs and a second slightly higher frequency, the phase quickly passes through an additional $-\pi$. The insets show this feature in more detail.}
    \label{fig:Ain phase plots}
\end{figure}

In Fig. \ref{fig:Ain phase plots}, we plot the phase $\phi(\tilde\omega)\equiv \text{Arg}(A_{\text{in}}(\tilde\omega))$ as a function of real frequency for the same combinations of $\delta M$ and $R$ for which we computed QNM frequencies. Note that because we defined $A_{\text{in}}$ with respect to a solution's internal frequency, whereas a distant observer measures a frequency of $\omega/\alpha$, the x-axis of these plots mark the observed frequency $\tilde\omega$ while the data at each $\tilde\omega$ is extracted for the internal frequency of that solution, which is $\tilde\omega\times\alpha$.

Previous work has shown that when a Schwarzschild BH is perturbed by some bump in the curvature potential, the local maxima of $|\phi'(\omega)|$ still align with the real part of the unperturbed QNM frequencies \cite{KyutokuQNMsRealAxis}.
When a bump is replaced by a shell, however, new subtleties emerge. We observe that across the majority of the real frequency domain, the evolution of $\phi(\tilde\omega)$ for the shell and vacuum systems is quite similar, and $|\phi'(\omega)|$ even retains a \textit{local} maximum near $\text{Re}(\omega_0)\approx 0.37/M$. However, at the real part of the weakly-damped QNM frequency identified in Sec. \ref{sec: polar qnm frequencies} and at a second slightly higher frequency, $A_{\text{in}}$ suddenly experiences a net phase change of $-\pi$, evolving much more rapidly than near $\omega_0$.

The net $-\pi$ phase change in $A_{\text{in,shell}}$ at these points agrees with the expected shift in $\phi(\omega)$ near the real part of \textit{vacuum} QNM frequencies given in \cite{KyutokuQNMsRealAxis}, but it actually emerges from localized scattering (e.g., from the shell) rather than scattering off the extended spacetime, as proposed in \cite{KyutokuQNMsRealAxis}. We previously connected the real part of the weakly-damped QNM with the resonant frequency of the shell's motion and the amplified excitation of secondary GWs. When a harmonic oscillator is driven above its resonant frequency, its motion is out of phase with the driving force. Similarly, when GWs incident on the shell have a frequency slightly above the resonant frequency, the shell's motion becomes out of phase with the incident wave. The secondary GWs also become out of phase, producing the phase shift of $-\pi$ in $A_\text{in}$.

The second jump of $-\pi$ in $\phi(\omega)$ also emerges from the amplified oscillations of the shell, but as we will see shortly, it actually corresponds to an \textit{anti}resonant feature in the secondary GWs. The total phase jump of $-2\pi$ ensures that at sufficiently low frequencies (i.e., at frequencies below the threshold where the additional mass $\delta M$ begins to substantially affect wave propagation in the exterior spacetime -- see the top plot in Fig. \ref{fig:Ain phase plots}), any changes to $\text{Arg}(A_\text{in})$ due to the shell are confined to a very narrow frequency band.

Despite the new sharp features in $\phi'(\omega)$, the connections proposed in \cite{KyutokuQNMsRealAxis} between the structure of $\phi(\omega)$ and ringdown stability are still reasonably valid, as the early ringdown remains dominated by the vacuum QNM frequencies. Nevertheless, the generation of new local maxima in $|\phi'(\omega)|$ by the shell suggests that there is space for further generalization of the conclusions presented in \cite{KyutokuQNMsRealAxis} to the broader space of possible perturbations to BH spacetimes.

\begin{figure}
    \centering
    \includegraphics[width=\linewidth]{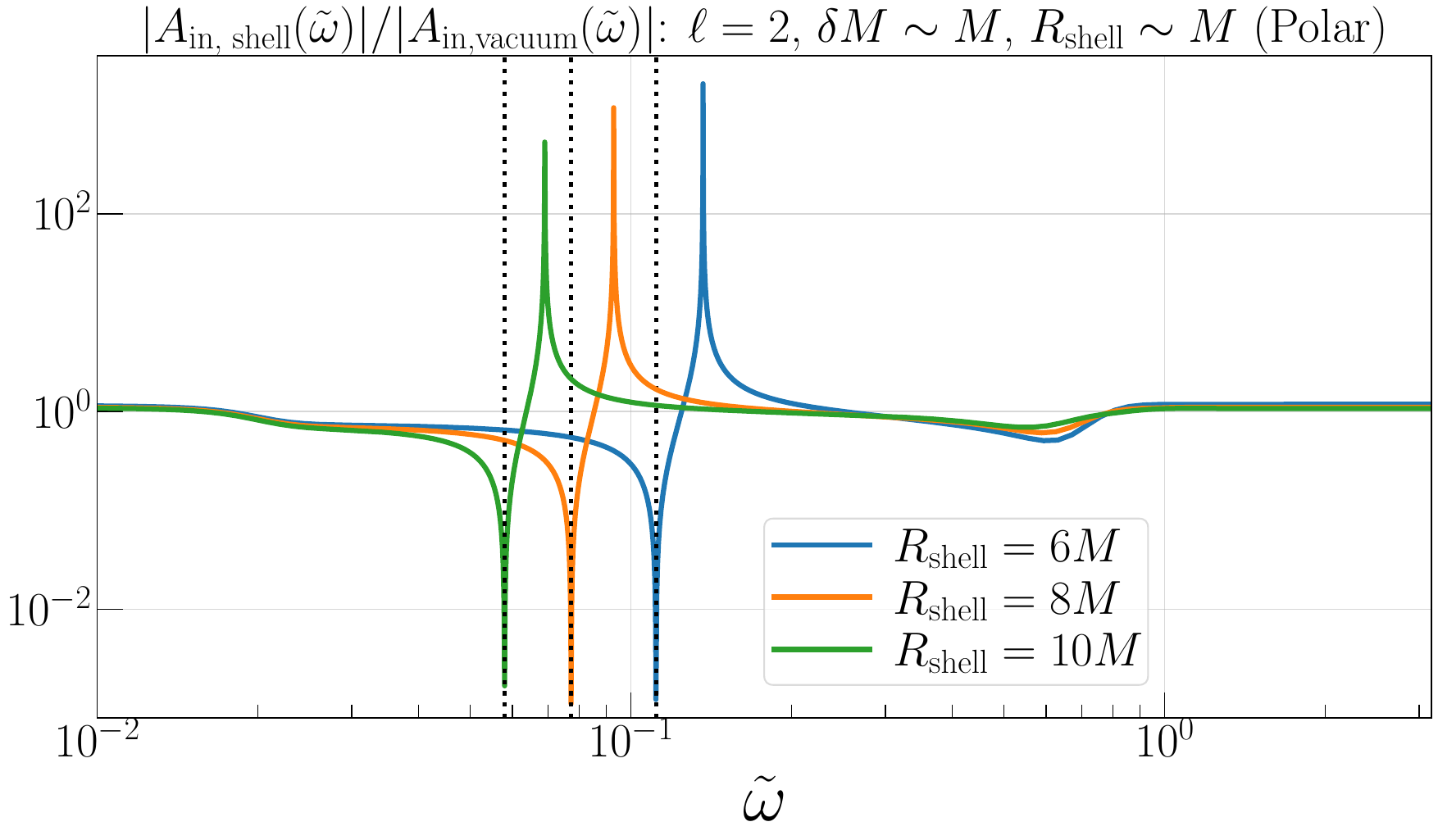}
    \includegraphics[width=\linewidth]{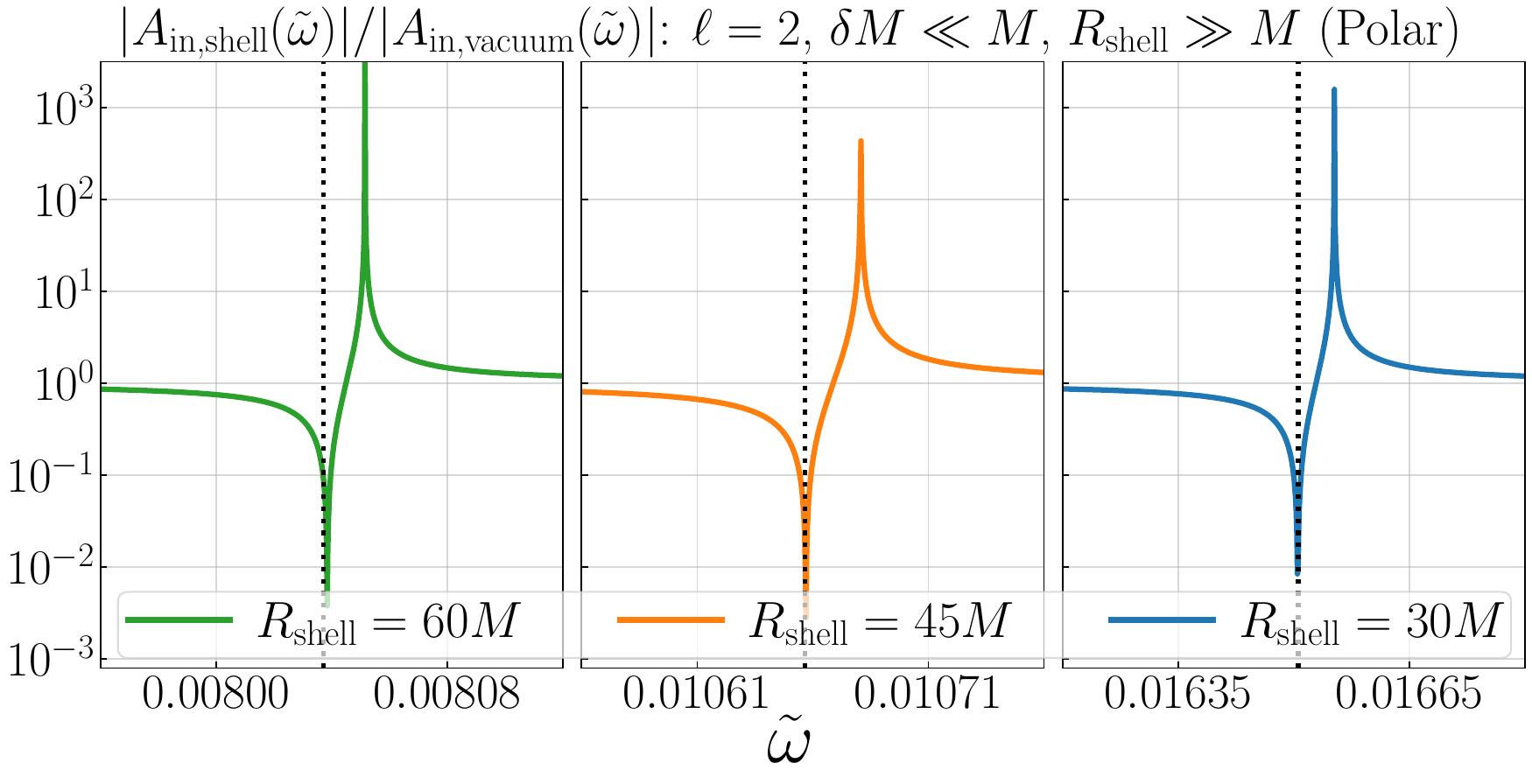}
    \caption{Top: the $\ell=2$ polar $|A_{\text{in,shell}}(\tilde\omega)/A_{\text{in,vacuum}}(\tilde\omega)|$ for a Schwarzschild with shell system with $\delta M=M/2$, $v_{\text{s}}=0.1$, and three choices of $R_{\text{shell}}\sim M$. Bottom: the same quantity, but with $\delta M=M/100$ and three choices of $R_{\text{shell}}\gg M$. Dotted black lines are placed at the real parts of the nearly-real QNM frequencies identified in Sec. \ref{sec: polar qnm frequencies} (divided by $\alpha$ to convert to observed frequency $\tilde\omega$). The behavior of $|A_{\text{in,shell}}|$ is only slightly modified relative to that of the vacuum system across the majority of real frequencies, but at the real part of the weakly-damped QNMs, this ratio reaches a local minimum followed by a pole at a slightly higher frequency. Furthermore, the full-width half-maximum of the resonant features matches the imaginary part of the respective weakly-damped QNM.}
    \label{fig:Ain magnitude plots}
\end{figure}

With evidence that the new QNMs influence the structure of $\text{Arg}(A_{\text{in}})$ within small real frequency bands, we then consider how the presence of the shell affects $|A_{\text{in}}|$, which we depict in Fig. \ref{fig:Ain magnitude plots}. Once again, $A_{\text{in}}$ is generally insensitive to the presence of the shell across the majority of the real frequency domain, especially in the case where $\delta M\ll M$ and $R_{\text{shell}}\gg M$. 

However, near the weakly-damped QNM frequencies identified in Sec. \ref{sec: polar qnm frequencies}, the ratio $|A_{\text{in,shell}}/A_{\text{in,vacuum}}|$ deviates significantly from unity. At the real part of the weakly-damped QNM frequency itself (black dotted lines in Fig. \ref{fig:Ain magnitude plots}), this ratio displays a local minimum; at the same slightly higher frequency where the second jump of $-\pi$ occurred in $\text{Arg}(A_\text{in})$, the ratio attains a sharp maximum. In fact, the QNMs associated with the weakly-damped shell ringing appear in the structure of $A_{\text{in}}(\omega)$ in the complex plane as a zero at the QNM frequency itself paired with a pole which lies on the real axis at a frequency just above the real part of the weakly-damped QNM. Interestingly, similar structure in $A_\text{in}$ on the real axis has been seen in studies of relativistic stars \cite{KojimaOscillationSpectra}.

We interpret the appearance of a paired zero and pole in $A_\text{in}(\omega)$ associated with a weakly-damped QNM as follows. Of course, the zero of $A_\text{in}$ is the QNM itself, where the real part of the mode is the resonant frequency of the shell's oscillations and the imaginary part is the inverse of the damping time due to gravitational radiation, which we saw scales with $\delta M$. At real frequencies slightly above the resonant frequency, the shell ringing is partially amplified and out of phase with the driving force, as is generally true for driven oscillator systems. The pole of $A_\text{in}$ occurs at the frequency where the partially amplified, out-of-phase secondary GWs destructively interfere with the primary excitation, so that the mode is completely suppressed at the horizon (i.e., $1/A_\text{in}\rightarrow0$). For real GW frequencies, energy conservation requires that $1+|A_\text{out}|^2=|A_\text{in}|^2$, with the resulting reflection and transmission coefficients (in power) of the spacetime being $|A_\text{out}/A_\text{in}|^2$ and $1/|A_\text{in}|^2$, respectively. Therefore, as $\omega$ approaches the pole frequency of $A_\text{in}$, $|A_\text{out}|$ must also become unbounded, and thus the composite Schwarzschild with shell spacetime is purely reflective to polar waves incident from infinity. Previous studies have found that a thin shell of matter can purely reflect waves of certain frequencies even when the interior metric is flat rather than our case of a Schwarzschild metric with smaller but nonzero mass \cite{Pfister_1996}.

Having seen that the real part of the weakly-damped QNMs determines the location of the sharp features on the real axis, we also observe that the imaginary part of the nearly-real frequencies is captured in the \textit{width} of these sharp features on the real axis. We are primarily interested in width of the features centered at local minima of $|A_\text{in,shell}|$, as these features are responsible for the amplified secondary GWs in ringdown waveforms. With this in mind, we have observed that when $|A_\text{in,shell}|$ attains a local minimum near the weakly-damped QNM frequency $\omega_{\text{Im}\ll\text{Re}}$, $\text{Im}(\omega_{\text{Im}\ll\text{Re}}/\alpha)$ matches the half-width half-maximum of the greybody factor $|1/A_{\text{in,shell}}|^2\equiv\Gamma_\text{shell}$. This behavior also follows in line with the general structure of mechanical resonances in the frequency domain.

In summary, the behavior of $|A_{\text{in,shell}}/A_{\text{in,vacuum}}|$ on the real frequency axis provides sufficient information to locate the complex QNM frequencies which lie near the real axis with decent accuracy. Once again, while we haven't presented the results explicitly here, we have confirmed that the same conclusions apply in cases where $\delta M\ll M$, $R_{\text{shell}}\sim M$ or $\delta M\sim M$, $R_{\text{shell}}\gg M$.

\subsection{Signatures of $A_{\text{in}}$ in Ringdown Waveforms}
\label{sec: Ain signatures in ringdown}
We observed in Sec. \ref{sec: ringdown fits} that for $\delta M\sim M$, $R_{\text{shell}}\sim M$, the late-time waveform features a clear signature of the shell's weakly-damped QNM, but for $\delta M\ll M$, $R_{\text{shell}}\gg M$, the weakly-damped QNM does not contribute substantially to the the late ringdown. Across all shell configurations we examined, the magnitude $A_\text{in,shell}$ at the shell resonant frequency is consistently smaller than its vacuum counterpart by two to three orders of magnitude. Therefore, the Fourier coefficient at the resonant frequency is two to three orders of magnitude larger in the shell case compared to the vacuum case, regardless of the shell parameters. As such, the stark difference in shell ringing amplitude must arise from other factors.

Since the time domain waveform is computed with an integral across the frequency domain with the integrand $\propto 1/A_\text{in}$, the width of the shell ringing features in $A_\text{in}$ on the real axis directly affects the amplitude of that component. For the sharp features in each configuration depicted in Fig. \ref{fig:Ain magnitude plots}, we have observed in the previous subsection that their widths are proportional to $\delta M$; furthermore, the local maximum value of $|1/A_{\text{in,shell}}|$ on the real axis near this feature stays roughly constant as $\delta M\rightarrow0$. After computing the inverse Fourier transform with an integral over this feature, the amplitude of this particular weakly-damped component therefore scales with $\delta M$, matching our observations in Sec. \ref{sec: ringdown fits}. 

However, we also observed that even when $\delta M\sim M$, if $R_\text{shell}$ is large, the shell ringing signal can still be quite weak (see Fig. \ref{fig:radial infall echo}). The strong dependence on the shell ringing amplitude on $R_\text{shell}$ lies in the behavior of $A_\text{in,vacuum}(\omega)$. Specifically, between frequencies of $\sim 0.01/M$ and $\sim 0.4/M$, $|A_\text{in,vacuum}|\sim\omega^{-4}$, while below $0.1/M$, $|A_\text{in,vacuum}|\sim \omega^{-9}$. So, even though $|A_\text{in,shell}/A_\text{in,vacuum}|$ can reach similar local minima regardless of the shell parameters, as the shell radius increases, its resonant frequency decreases (see \ref{sec: polar qnms interpretation}) so that $|A_\text{in,vacuum}|$ becomes many orders of magnitude larger at the resonant frequency. Despite its resonant amplification, the shell ringing mode is still highly suppressed as $R_\text{shell}$ grows, producing the hierarchy observed in the ringdown waveforms. In principle, though, unlimited numerical precision would enable observation of the shell ringing in any ringdown waveform once enough time has passed for the initial ringdown, echoes, and tail to decay below the level of the ringing signal.

The pole in $A_\text{in,shell}$ just above the resonant frequency does not significantly impact ringdown waveforms. Since this mode is highly suppressed when the shell is present, its contribution is limited by the magnitude of the corresponding Fourier component in vacuum\footnote{Perfectly suppressing a mode is equivalent to adding in a component in the frequency domain with an amplitude mirroring that of the corresponding component in the unperturbed waveform. Since the vacuum ringdown does not feature a weakly-damped ringing contribution, the suppression at the real pole frequency actually adds in a long-lived contribution, but its size is limited by the Fourier transform of the vacuum ringdown waveform at that frequency and the width of the feature.}. Therefore, the amplified component from the resonant frequency itself always dominates the total shell ringing signal.

\subsection{$A_{\text{in}}$ in the Axial Sector}
\begin{figure}
    \centering
    \includegraphics[width=\linewidth]{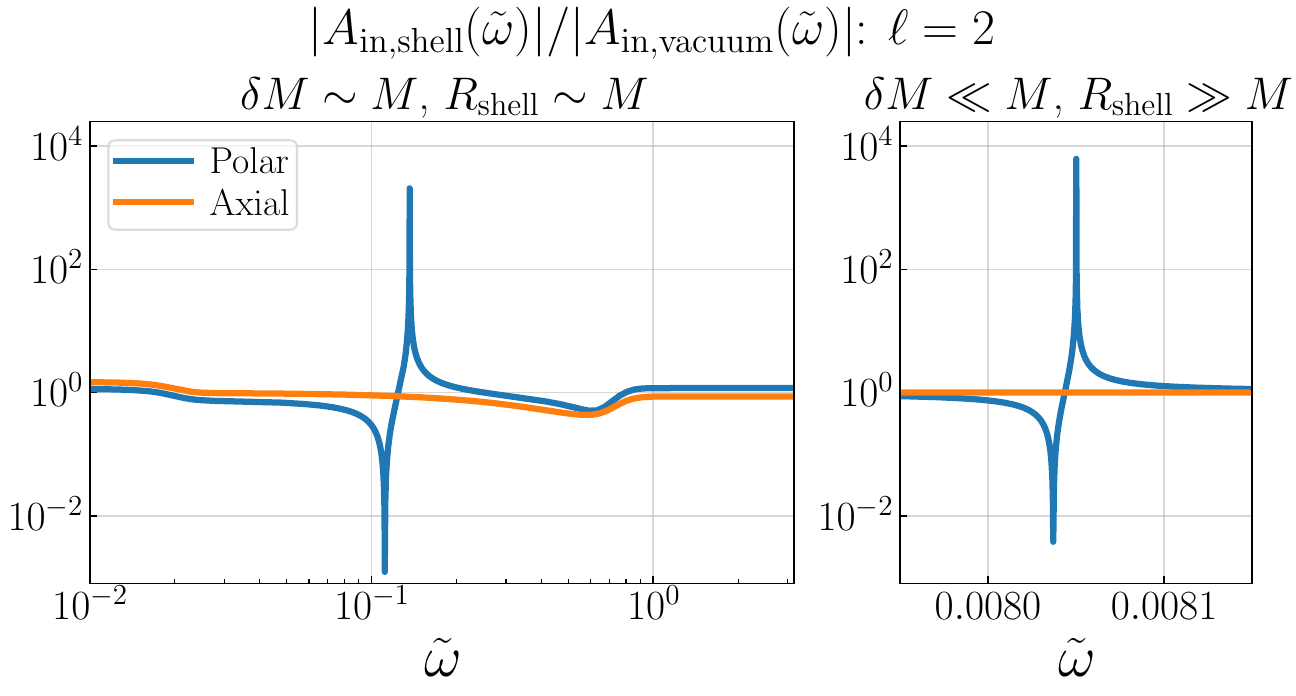}
    \caption{The $\ell=2$ polar and axial $|A_{\text{in}}(\omega)|$ for the Schwarzschild with shell spacetime. Top: $\delta M=M/2$, $v_{\text{s}}=0.1$, $R_{\text{shell}}=6M$. Bottom: $\delta M=M/100$, $v_{\text{s}}=0.1$, $R_{\text{shell}}=30M$. Since axial metric perturbations cannot produce secondary GWs from the spherically-symmetric shell, the axial $A_{\text{in}}$ does not demonstrate sharp features when the shell is added, unlike its polar analogue. In the case where $\delta M\sim M$ and $R_{\text{shell}}\sim M$, the axial $A_{\text{in,shell}}$ only deviates from the vacuum case due to the additional mass affecting wave propagation substantially in the exterior region.}
    \label{fig:Ain magnitude polar vs axial}
\end{figure}

The lack of coupling between axial perturbations and the shell matter is especially apparent in the structure of the axial $A_{\text{in}}$ on the real axis. In Fig. \ref{fig:Ain magnitude polar vs axial}, we present the behavior of the polar and axial $|A_{\text{in}}|$ for two representative shell configurations. Immediately, we observe that the axial sector does not feature the modes of amplified excitation relative to the vacuum case which appear in polar sector, consistent with the lack of nearly-real axial QNM frequencies. It follows then that just as we observed in the polar sector, the shell should source purely perturbative changes to axial ringdown waveforms for sufficiently small $\delta M$, regardless of the shell radius. Thus, the early ringdown of the Schwarzschild with shell system is stable in both the polar and axial sectors.

Previous studies examining the impact of spherically-symmetric external mass distributions on ringdown signals have generally been limited to axial perturbations, though more recent work has started to fill the gaps in the polar sector \cite{SpeeneyPolarPerturbations}. Our toy model emphasizes the importance of studying polar modes in spherically-symmetric systems, as the differences in the GW-matter coupling between the polar and axial sector enable the polar modes to carry traces of physical phenomena which axial modes are inherently insensitive to.

\subsection{Summary of $A_\text{in}$ Features}
We now provide a brief summary of our findings regarding the impact of the shell on $A_\text{in}$.
\begin{enumerate}[i.]
    \item The phase of $A_\text{in}$ generally evolves in the same way as the vacuum case, except near the real part of the weakly-damped QNM frequencies, where it rapidly gains a phase of $-\pi$ relative to the vacuum case at both the resonant frequency and a second slightly higher frequency.
    \item At the shell resonant frequency, the magnitude of $A_\text{in}$ is orders of magnitude smaller than the vacuum case -- this feature corresponds to the amplified secondary GWs at late times. The effective full-width half-maximum of $1/|A_\text{in}|^2$ matches imaginary part of the QNM frequency. Since the width scales with $\delta M$, the shell ringing amplitude also scales with $\delta M$, consistent with the picture of secondary quadrupole radiation generated by the shell. However, the shell ringing expresses a much stronger dependence on $R_\text{shell}$, originating from the shape of $A_\text{in,vacuum}(\omega)$.
    \item A pole in $A_\text{in}$ lies on the real axis just above the real part of the weakly-damped QNM frequencies; it physically emerges from the destructive interference between primary excitations and the secondary GWs produced by the out of phase shell ringing.
    \item In the axial sector, the lack of coupling between the metric perturbations and shell oscillations ensures that for $\delta M\ll M$, the behavior of $A_\text{in}$ on the real axis does not differ significantly from the vacuum case for sufficiently low frequencies.
\end{enumerate}

\section{Ringdown Waveforms and the Full and Rational Filters}
\label{sec: ringdown filters}
An alternative method to fitting the QNM frequencies for parameterizing the ringdown of a compact object is through the ringdown filters presented in \cite{SizhengFilters}. The authors present two classes of filters which in principle should remove the ringdown portion from a frequency domain waveform. The first is the rational filter, given by the following product over QNM frequencies $\omega_i$: 
\begin{equation}
    \mathcal{F}_{\text{rat}}(\tilde\omega)=\prod_i\frac{\tilde\omega-\omega_{i}/\alpha}{\tilde\omega-\omega_{i}^*/\alpha}.
\end{equation}
The filter is normalized such that $|\mathcal{F}_{\text{rat}}|=1$, which ensures that it does not change the total power in the waveform.
For a simple single-QNM ringdown $\mathcal{Z}(t)=\exp(i\omega_{1}t)\Theta(t)$, where $\Theta$ is the Heaviside step function, the rationally-filtered waveform is $\mathcal{Z}_{\text{filter}}(t)=-\exp(-i\omega_{1}^*t)\Theta(-t)$, removing the ringdown feature for $t>0$ and converting it to a ring-up feature for $t<0$. In other words, the QNM is reflected across its start time and multiplied by $-1$. A waveform constructed purely from a sum of decaying sinusoids with fixed frequency and decay time vanishes for times after the start of the ringdown after the rational filter product is applied.

The authors of \cite{SizhengFilters} also propose that $A_{\text{in}}(\omega)$ in the vicinity of QNM frequencies should behave roughly as\footnote{This same proposed form is used by the authors of \cite{KyutokuQNMsRealAxis} to explain the features of $\phi(\tilde\omega)$ that we explored in Sec. \ref{sec: Ain on real axis}.} a product of $(\omega-\omega_\text{QNM})$ with some slowly-varying function of $\omega$. With this in hand, they construct the second class of filter, called the ``full filter", which can defined by
\begin{equation}
    \mathcal{F}(\tilde\omega)\equiv\frac{A_{\text{in}}(\tilde\omega)}{A_{\text{in}}^*(\tilde\omega)}=\exp(2i\phi(\tilde\omega)),~\phi\equiv \text{Arg}(A_{\text{in}}).
\end{equation}
Application of both the rational and full filters to compact merger numerical relativity waveforms can quite effectively filter out the quasinormal behavior from the post-merger signal \cite{SizhengFilters}. 

Note that we have found that $A_\text{in}$ features a pole at positive integer multiples of the ``horizon mode" frequency $\omega_H=-i/4M$ \cite{ZimmermanHorizonMode} in both the vacuum and shell systems. Since the full filter removes QNM frequencies, with $A_\text{in}$ behaving roughly as $(\omega-\omega_\text{QNM})$, it actually \textit{adds} in a contribution at the horizon mode frequency, with $A_\text{in}$ behaving roughly as $(\omega-\omega_\text{H})^{-1}$. While we were not able to conclusively determine whether these horizon modes (or ``redshift modes", in the terminology of \cite{DeAmicis:2025xuh}, for example) appear in the unfiltered waveforms, when applying the full filters in this work, we combine them with a rational filter at $\omega_H$ to remove this artificially generated component.
In principle, the full filter applied in \cite{SizhengFilters}, which is constructed from $D_\text{out}$, the outgoing wave amplitude at the horizon for the ``up" solution to the Teukolsky radial equation, does not introduce this horizon mode feature. In vacuum, the Regge-Wheeler and Zerilli amplitudes $A_\text{in}$ are related to $D_\text{out}$ by well-known transformations \cite{SasakiReview}, but we leave the extension of such transformations to account for the effects of the shell to future work.

While the filter is at least nominally connected to the greybody factor through the appearance of $A_{\text{in}}$, these two quantities \textit{a priori} capture two unique sets of information, with the filter encoding the phase of $A_{\text{in}}$ and the greybody factor depending only on its magnitude. That being said, we have seen in the previous section that the magnitude and phase of $A_\text{in}$ deviate from their vacuum values in similar frequency bands -- the same shell physics is encoded in the structure of both quantities.

The structure of the full and rational filters are uniquely defined by the properties of the BH alone (in the vacuum case) or the BH and shell (in the shell case). With that in mind, if either the full or rational filters can convincingly suppress the ringdown portion of a general merger waveform, identifying the most effective filter for given compact merger GW data could offer a reliable alternative method beyond extracting QNM frequencies and amplitudes for inferring the properties of the remnant and its local environment. 

To test the viability of this BH filter program when applied to more general astrophysical environments, we now wish to evaluate the performance of the full and rational filters, constructed in the shell and vacuum models, in removing the ringdown signature from our radial infall waveforms. To that end, we define the following score quantity to compare the effectiveness of each filter:
\begin{equation}
    Y_j(t_1,t_2)\equiv1-\int_{t_\text{peak}+t_1}^{t_\text{peak}+t_2}\frac{|\mathcal{H}(\mathcal{Z}_j(t))/\mathcal{H}(\mathcal{Z}_{\text{no filter}}(t))|}{t_2-t_1}dt,
\end{equation}
where $\mathcal{Z}_j$ is a filtered waveform, given by
\begin{equation}
    \mathcal{Z}_j(t)=\int_{-\infty}^{\infty}\mathcal{Z}(\tilde\omega)\mathcal{F}_j(\tilde\omega)e^{-i\tilde\omega t} d\tilde\omega,
\end{equation}
with $j\in\{$no filter, vacuum rational, vacuum full, shell rational, shell full$\}$, and $t_\text{peak}$ again is the time at which $|\mathcal{Z}_{\text{no filter}}(t)|$ is maximized. Note that since $\mathcal{Z}(t)$ is real, we must apply the rational filters with both $\omega_i=\omega_\text{QNM}$ and $\omega_i=-\bar\omega_\text{QNM}$ to properly filter that QNM frequency. The Hilbert transform $\mathcal{H}(\mathcal{Z}(t))$ converts the purely-real $\mathcal{Z}(t)$ into a complex signal whose norm corresponds to the amplitude envelope of the waveform. The score is normalized so that $Y_{\text{no filter}}=0$, and if a filter cleanly removes the ringdown features after peak strain, its corresponding value $Y_j$ should be near unity. 

\begin{figure}[!ht]
    \centering
    \includegraphics[width=\linewidth]{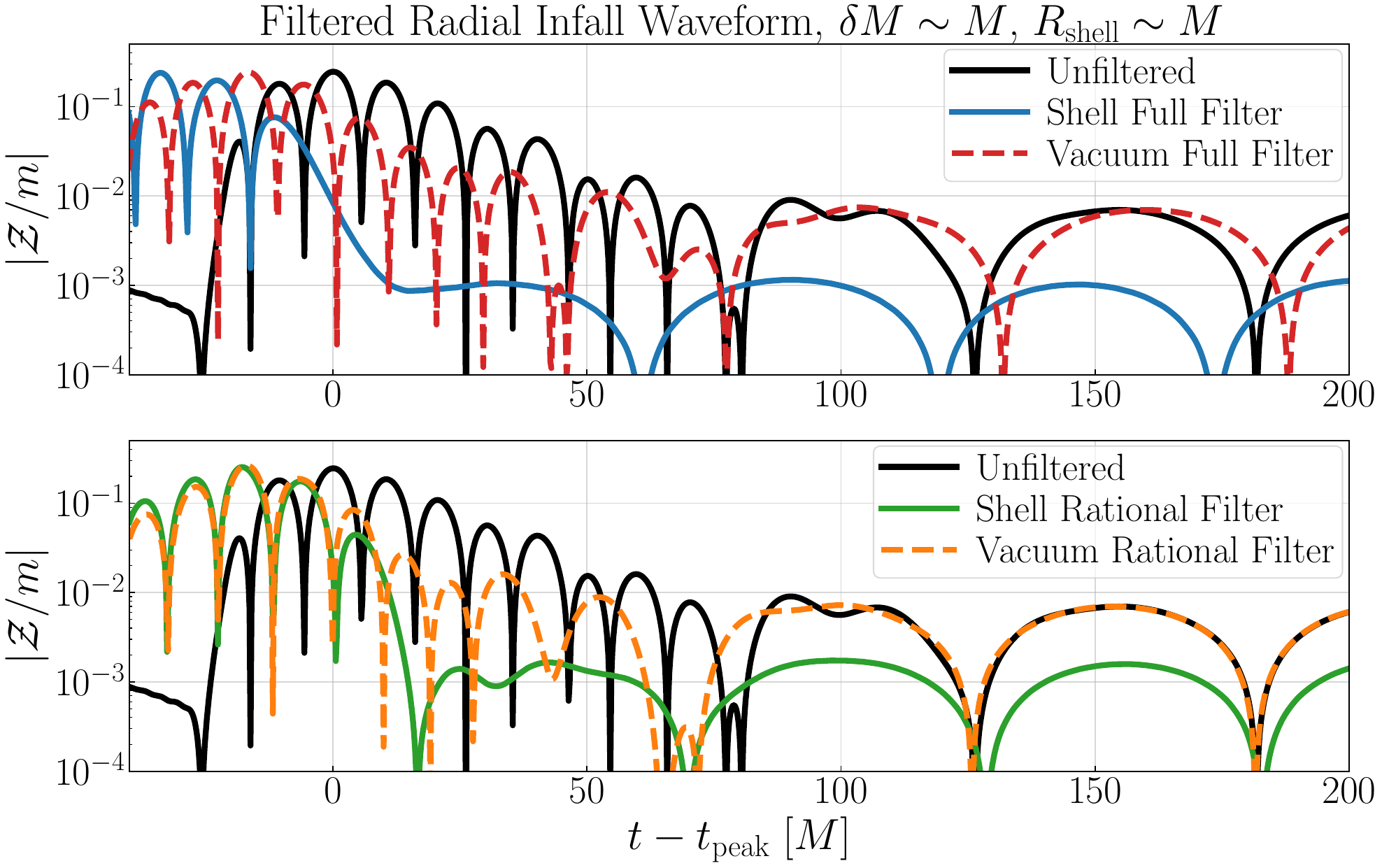}
    \\
    \vspace*{0.1cm}
    \includegraphics[width=\linewidth]{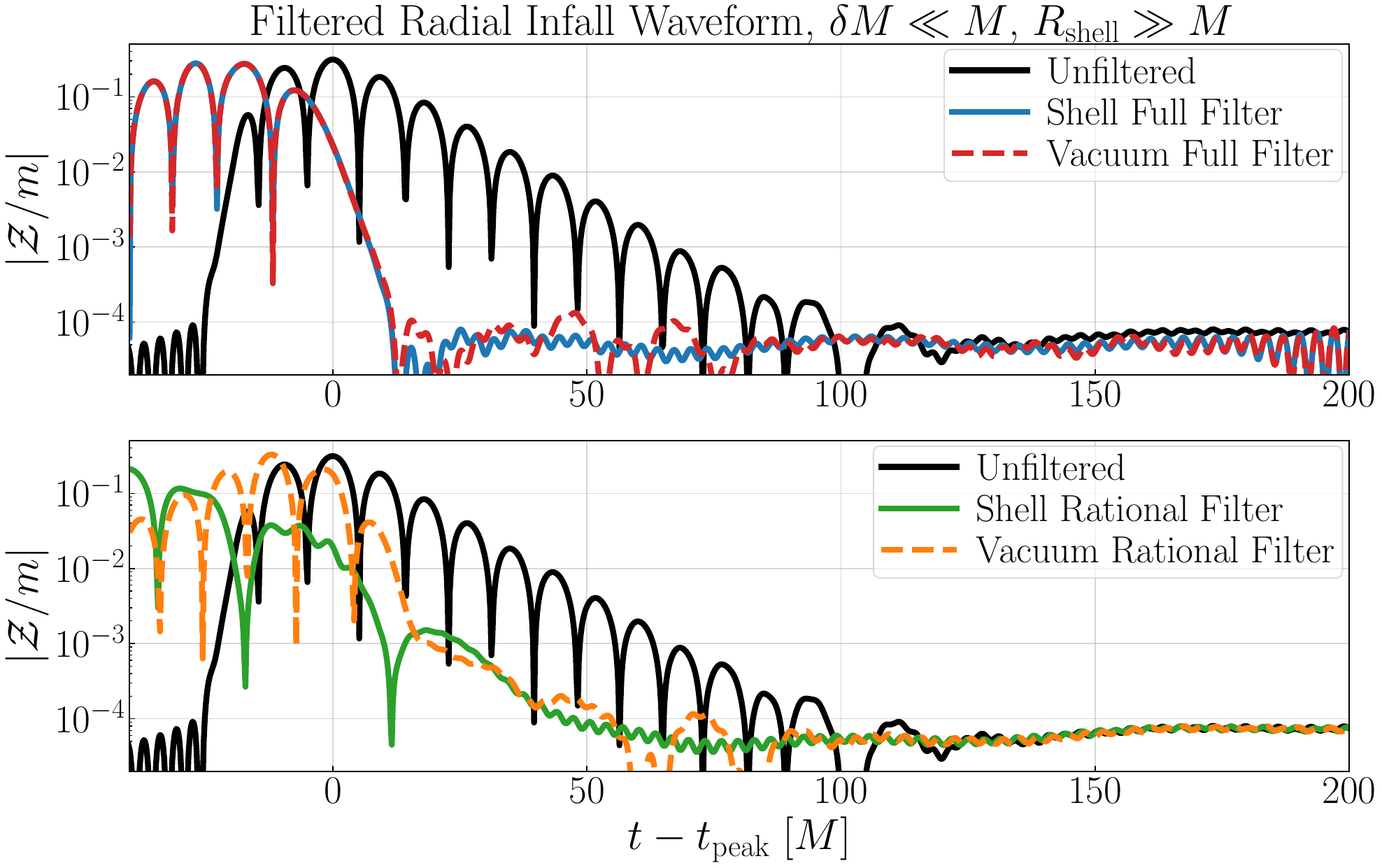}
    \\
    \vspace*{0.1cm}
    \includegraphics[width=\linewidth]{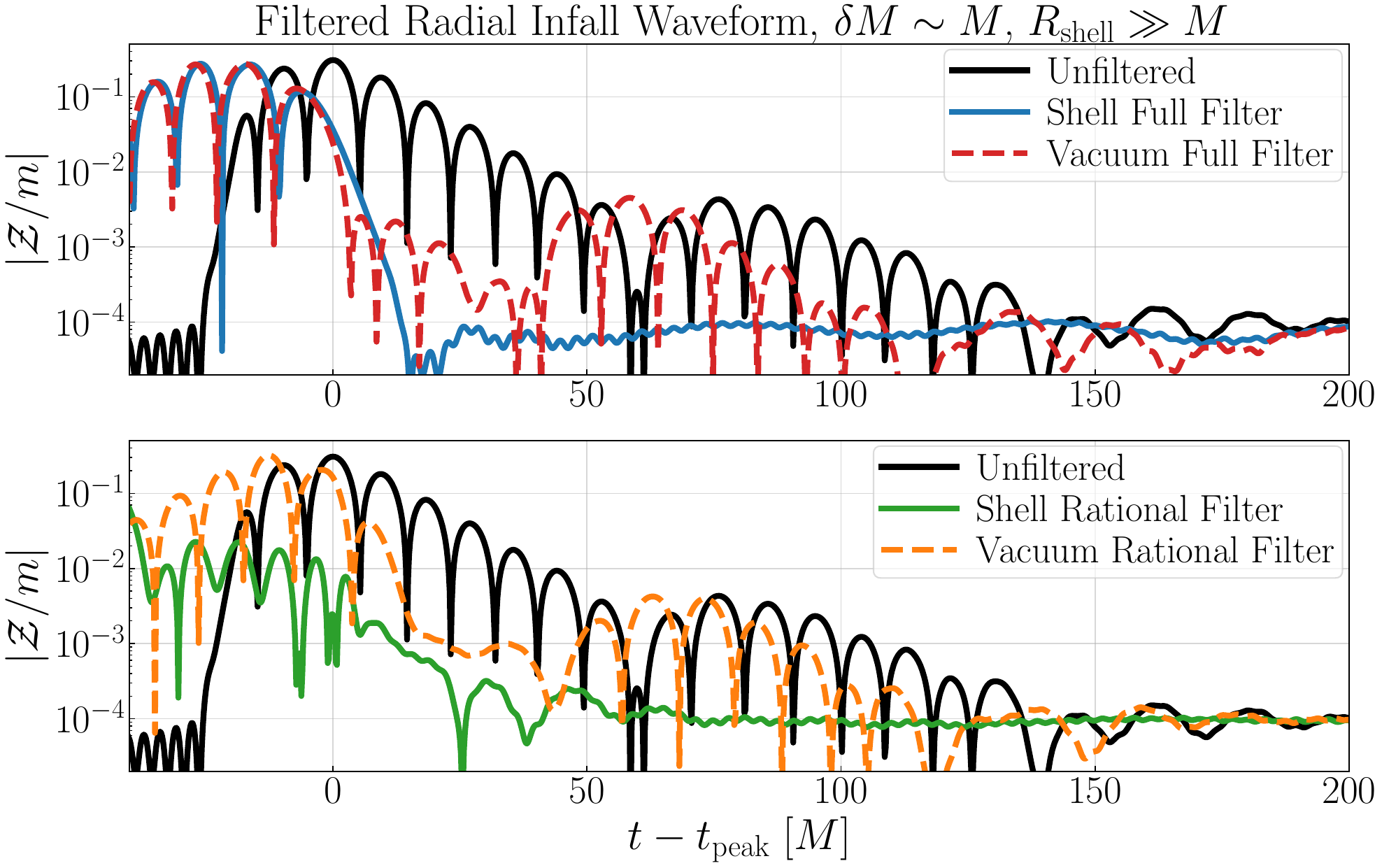}
    \caption{Radial-infall waveforms for $\ell=2$ in the Schwarzschild with shell spacetime along with the filtered waveforms after application of the shell and the vacuum full and rational filters. Top plots: $\delta M=M/2$, $v_{\text{s}}=0.1$, $R_{\text{shell}}=6M$. Middle plots: $\delta M=M/100$, $v_{\text{s}}=0.1$, $R_\text{{shell}}=30M$. Bottom plots: $\delta M=M/2$, $v_{\text{s}}=0.1$, $R_\text{{shell}}=30M$. In the case where $\delta M\sim M$, $R_{\text{shell}}\sim M$, the shell filters are more effective at removing the weakly-damped shell ringing. Interestingly, at early times in the $\delta M\ll M$, $R_\text{shell}\gg M$ case, the full filters act almost identically on the early ringdown. Also, when an echo is present, the shell filters appear far more effective than their vacuum counterparts at removing it from the waveform.}
    \label{fig:radial infall waveform filter tests}
\end{figure}

In Fig. \ref{fig:radial infall waveform filter tests}, we present the same three radial infall waveforms that were computed in Sec. \ref{sec: ringdown fits}, along with the waveforms after application of each test filter. In computing the rational filters, we use four QNMs for the vacuum case and the nine modes with smallest $|\omega_I|$ for the shell case\footnote{Only two QNMs -- a migrating mode and a weakly-damped mode -- were found for the case with $\delta M=M/2$, $R_\text{shell}=6M$, so those are the only modes used for that configuration.}. The scores $Y_j$ for each filter, evaluated for the time intervals of $[0,50]M$ and $[0,150]M$ after peak strain, appear in Table \ref{tab:filter test results}.

\begin{table}[t]
    \centering
    Filter Scores $Y_j$ for Radial Infall Waveform
    \\
    \vspace*{0.1cm}
    \begin{tabular}{|c|c|c|c|c|c|c|}
         \hline
          & \multicolumn{2}{|c|}{Shell 1} & \multicolumn{2}{|c|}{Shell 2}
          & \multicolumn{2}{|c|}{Shell 3}
         \\
         \hline
         \textbf{Filter} & A & B & A & B & A & B
         \\
         \hline
         Vac. rtnl. & 0.7689 & 0.2595  & 0.8852  & 0.3926 & 0.8752 & 0.1442
         \\
         \hline
         Vac. full & 0.7204  & 0.2341 & 0.9404  & 0.4930 & 0.8878 & 0.1712
         \\
         \hline
         Shell rtnl. & 0.9251 & 0.8180 & 0.9486 & 0.4276 & 0.9506 & 0.6862
         \\
         \hline  
         Shell full & 0.9679 & 0.8856 & 0.9401 & 0.4957 & 0.9353 & 0.6998
         \\
         \hline
    \end{tabular}
    \caption{Filter scores $Y_j(t_1,t_2)$ for the radial infall ringdown waveforms for three shells (all with $v_\text{s}=0.1$): Shell 1 ($\delta M=M/2$, $R=6M$), Shell 2 ($\delta M=M/100$, $R=30M$), and Shell 3 ($\delta M=M/2$, $R=30M$). We test the rational (rtnl.) filter with vacuum (vac.) QNM frequencies, the full filter for the vacuum spacetime, the rational filter with shell QNM frequencies, and the full filter for the shell spacetime. 
    Filter scores are calculated for Window A, with $(t_1,t_2)=(0,50M)$ and Window B, with $(t_1,t_2)=(0,150M)$. 
    }
    \label{tab:filter test results}
\end{table}

The filter performances vary primarily in their ability to subtract the late-time shell signatures from the waveform. Indeed, in the case where $\delta M\sim M$, $R_{\text{shell}}\sim M$, the shell filters substantially outperform the vacuum filters both in the early and late ringdown, as the shell ringing is strong in this system and thus becomes a significant component of the total ringdown signal at early times. Visual examination of the filtered waveforms in Fig. \ref{fig:radial infall waveform filter tests} confirms that application of the vacuum filters leaves the shell ringing intact (up to a phase shift) while the shell filters attenuate the weakly damped ringing. Furthermore, the shell full filter also outperforms the shell rational filter, both in the early and late times. 

Curiously, the shell filters do not completely remove the shell ringing feature, leaving behind a smaller weakly-damped component with frequency slightly higher than the primary shell ringing component. In the case of the rational filter, we suspect this emerges from the highly suppressed mode at the real pole frequency of $A_\text{in,shell}$ (see Sec. \ref{sec: Ain on real axis}), which is not a QNM frequency and thus not filtered out. Meanwhile, much like the pole in $A_\text{in}$ at the horizon frequency introduces a component with frequency $\omega_H$ when the full filter is applied, the pole in $A_\text{in,shell}$ near the resonance also introduces a weak component after application of the shell full filter.

For the case where the shell ringing is much weaker ($\delta M\ll M$, $R_{\text{shell}}\gg M$), however, the vacuum and shell filters are not as obviously distinguished. The filter scores imply that all four filters perform comparably to one another throughout ringdown, with the vacuum rational filter only slightly worse than the others. At later times, the general performance of the filters becomes worse, but this is likely an artifact of the numerical noise floor limiting the effectiveness of the filters as the waveform decays. The shell and vacuum full filters produce filtered waveforms which appear nearly visually identical. This suggests that certain features we observed in Sec.~\ref{sec: ringdown fits} -- namely, the ability of vacuum QNMs to fit early shell ringdown waveforms with $\delta M \ll M$ and $R_{\text{shell}} \gg M$, and of the destabilizing shell modes to reproduce vacuum ringdown waveforms at early times -- are encoded in the structure of the full filter itself.

Finally, in the case where the shell produces an observable echo, the shell filters clearly outperform their vacuum counterparts. After application of the shell full filter to the unfiltered ringdown, the resulting waveform looks strikingly similar to that of the shell system with $\delta M\ll M$, $R_\text{shell}\gg M$, which did not feature an echo. This implies that the structure of the echo may also encoded directly within $A_\text{in,shell}$ -- such behavior has been previously observed for echo-producing exotic compact objects \cite{SizhengFilters}. For now, we leave the identification of these properties for future work.

Given that the rational filter reflects a decaying sinusoid across its starting time, we expect that since the echoes have a starting time well after $t_\text{peak}$, the vacuum rational filter (and by extension, the vacuum full filter, given the form it takes near the QNM frequencies of $(\omega-\omega_\text{QNM})$ $\times$ some slowly varying function) should contaminate the early ringdown with the reflected echo component. Indeed, we see that both vacuum filters produce a ring-up feature just before the start of the echo.

\begin{figure}[t]
    \centering
    \includegraphics[width=\linewidth]{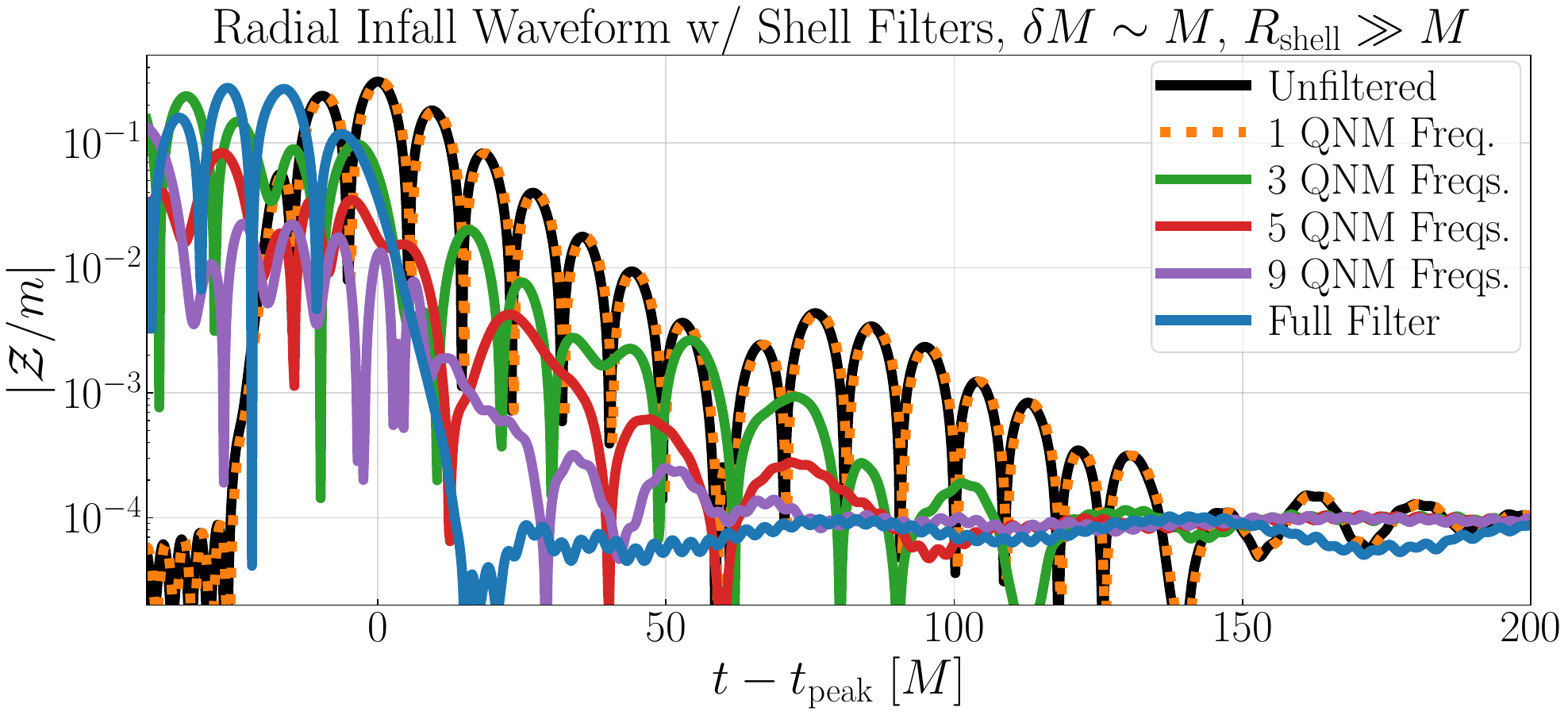}
    \caption{The $\ell=2$ radial-infall waveform with $\delta M=M/2$, $v_{\text{s}}=0.1$, $R_\text{{shell}}=30M$, along with the results after application of the shell rational filters with various numbers of modes and the shell full filter.
    Modes are added to the rational filter in order of increasing $-\omega_I$. As more modes are added, both the early ringdown and echo are gradually removed from the waveform for $t>t_\text{peak}$. The first filter creates very little distinction because it is built from the shell ringing mode, which is not excited much in this system due to its low frequency.}
    \label{fig:radial infall waveform rational filter progression}
\end{figure}

Interestingly, the shell rational filter does not appear to leave behind this ring-up component. However, we suspect this behavior simply emerges from the many closely-spaced QNM frequencies which appear when the spectrum is destabilized. We saw in Sec. \ref{sec: ringdown fits} that the echo can be fit with just a few destabilizing mode frequencies; after applying many individual rational filters whose frequencies lie near those of the QNMs which fit the echo (see the blue squares in Fig. \ref{fig:destabilized spectrum large radii}), the echo signal may be shifted backwards multiple times, sending it to times before $t_\text{peak}$. In fact, if we reduce the number of shell QNMs which are used in the rational filter, contaminants appear in the early ringdown as expected (see Fig. \ref{fig:radial infall waveform rational filter progression}). This behavior may also explain the efficacy of the shell full filter, as it in principle naturally carries a factor of $(\omega-\omega_\text{QNM})$ for each mode in the destabilized spectrum.

Ultimately, we conclude that distinctions between the performances of vacuum and shell filters only become significant when features which cannot appear in vacuum dominate the GW strain. Based on the ringdown waveforms we have produced here, it is evident that these effects would only appear in true data streams either with \textit{i)} substantially-improved detector sensitivities, allowing the late ringdown to be observed, or \textit{ii)} a merger occurring in an environment surrounded by a shell of mass comparable to the remnant BH (which we suspect is likely quite rare). Out of all the shell configurations considered, the only ones which grant the filters a competitive advantage over simple QNM fits in data analyses are those which produce a strong echo, as the shell filters are apparently able to naturally remove that late-time feature. However, this comes with the caveat that a rational filter composed from a tightly-spaced collection of destabilizing modes can push an echo to times far before $t_\text{peak}$, an effect that the widely-spaced vacuum QNMs cannot achieve. Therefore, more careful analysis of the properties of these filters is needed in the future before they can be genuinely applied to data streams to search for echo signatures.

Our discussion of the shell and vacuum filter performances, especially in handling late-time shell features, hints at just some of the systematic limitations to the ringdown filter program that may be imposed by general environments around remnant black holes. A thorough treatment on such systematic issues will be an important step in understanding the capabilities of the ringdown filters to extract remnant and environmental parameters from ringdown waveforms.

\section{Conclusion and Future Directions}
\label{sec: conclusion}
In this paper, we studied the effect of a thin shell of matter surrounding a Schwarzschild BH on the ringdown properties of the system. Like previous studies which have considered localized perturbations to BH exteriors using ``bumps" in the curvature potential, we found that when the shell mass is sufficiently small, the fundamental mode migrates perturbatively, but when the shell is placed far from the central BH ($R_{\text{shell}}\gg M$), the QNM spectrum is destabilized, with many new modes appearing in proximity to the original fundamental mode. Unlike these previous works, however, we find that the shell generates an additional weakly-damped QNM in the polar sector only; physically, this mode arises from the secondary GWs sourced by the resonantly-amplified motion of the shell which is driven by its coupling to the primary metric perturbations. These nearly-real QNMs generate sharp features in $A_{\text{in,shell}}(\omega)/A_\text{in,vacuum}(\omega)$ on the real axis -- a feature not previously observed for models with localized perturbations to the curvature potential. Furthermore, these features carry signatures of the physical interactions between metric perturbations and the shell and translate effectively to behavior observed in ringdown waveforms.  

Notably, the inability of axial perturbations to excite secondary GWs from the shell results in \textit{i)} the lack of nearly-real QNM frequencies in this sector, and \textit{ii)} the insensitivity of the axial $A_{\text{in}}$ to the addition of a shell.

We considered the role of the shell on least-squares fits of ringdown waveforms with QNMs and found that the shell system QNMs provide a significantly better mismatch when the shell ringing is strong ($\delta M\sim M$, $R_\text{shell}\sim M$) or when an echo appears ($\delta M\sim M$, $R_\text{shell}\gg M$), as the early ringdown is still well described by the vacuum system QNMs. We also considered the full and rational ringdown filters proposed in \cite{SizhengFilters}. Our results after application of these filters to the ringdown waveforms in the shell system mirror those of our tests with QNM fitting. Namely, the performance of the shell filters exceeds that of the vacuum filters only when signatures exclusive to the shell, such as the weakly-damped QNMs, become significant, as the shell filters can handle such features while the vacuum filters cannot. When the shell sources an echo, the shell filters also outperform their vacuum counterparts, but more analysis is needed to fully understand whether this advantage will be useful for searching for echoes in GW observatory data.

This work highlights some of the many subtleties that must be addressed in accurately modeling astrophysical environments which surround GW sources, such as active galactic nuclei, accretion disks, and clouds of scalar or vector fields. Studies of more extended matter distributions have also uncovered evidence of late-time matter oscillations in the polar sector \cite{CardosoHaloPerturbationFormulas}; it could be interesting to examine how the structures in $A_\text{in,shell}$ evolve as the shell thickness increases beyond the infinitesimal limit studied in this work. On a similar note, one might more thoroughly probe the effects of matter coupling on GW propagation by giving the shell some non-gravitational damping mechanism and studying its impact on the shell QNM frequencies and behavior of $A_\text{in}$. 

Of course, extending this study to the rotating case is a worthwhile direction given that most remnant BHs observed in data have substantial spin. Furthermore, beginning the test particle's radial infall trajectory outside the shell (instead of inside, as we did here) would create an offset between the start times of the GW signals from the infalling particle and secondary shell ringing, which might impact the performance of the full and rational ringdown filters. Such future directions will work towards filling gaps in our understanding of how local astrophysical environments affect GW signals, enabling more accurate data analysis with the next-generation GW observatories scheduled for construction in the coming decades.

\begin{acknowledgements}
A.L. is grateful for support from the Fannie and John Hertz Foundation. D.L. acknowledges support from the Simons Foundation through Award No. 896696, NSF Grant No. PHY-2207650, and the National Aeronautics and Space Administration through award 80NSSC22K0806. C.W. and Y.C.’s research is supported by the Simons Foundation (Award No. 568762), the Brinson Foundation, and the National Science Foundation (via Grants No.  PHY-2309211 and No. PHY 2309231). We thank Brian Seymour, Adrian Ka-Wai Chung, Sizheng Ma, Isaac Legred, and Ling Sun for helpful discussions.
\end{acknowledgements}

\appendix

\section{More on the Junction Conditions}
\label{app: junction conditions}
In Sec. \ref{sec: junction conditions}, we outlined our procedure for incorporating the effect of the stress-energy of the thin shell of matter into the global homogeneous solutions to the Zerilli and Regge-Wheeler equations. We now provide some further clarification on the mathematical structure present in the polar version this procedure.

In the polar sector, we must use all four junction conditions listed in Eq. \eqref{eq:RW gauge jumps}, for even though the goal of constructing the global homogeneous solutions only requires finding $[[Z]]\equiv Z_\text{ext}(R)-Z_\text{int}(R)$ and $[[Z']]\equiv Z'_\text{ext}(R)-Z'_\text{int}(R)$, the fixed-worldtube gauge variable $z$ and energy density perturbation $\delta \Sigma$ are additional unknowns which depend on the local metric perturbation and thus on $Z$ and $Z'$.

After using Eq. \eqref{eq: polar conversions} and the Zerilli equation to rewrite all instances of $K$, $K'$, $H$, $H'$, and $H_1$ in Eq. \eqref{eq:RW gauge jumps} in terms of $Z$ and $Z'$, making sure to account for the increased exterior mass $M+\delta M$ and reduced exterior frequency $\omega/\alpha$, this system of four equations can be expressed as a simple linear matrix equation. In other words, we find
\begin{multline}
\label{eq: polar matrix equation}
    \begin{pmatrix}
        A_{11} & A_{12} & A_{13} & A_{14}
        \\
        A_{21} & A_{22} & A_{23} & A_{24}
        \\
        A_{31} & A_{32} & A_{33} & A_{34}
        \\
        A_{41} & A_{42} & A_{43} & A_{44}
    \end{pmatrix}
    \begin{pmatrix}
        Z_\text{ext}(R)
        \\
        Z'_\text{ext}(R)
        \\
        z
        \\
        \delta\Sigma
    \end{pmatrix}
    \\
    =
    \begin{pmatrix}
        B_1
        \\
        B_2
        \\
        B_3
        \\
        B_4
    \end{pmatrix}
    Z_\text{int}(R)
    +
    \begin{pmatrix}
        C_1
        \\
        C_2
        \\
        C_3
        \\
        C_4
    \end{pmatrix}
    Z'_\text{int}(R)
\end{multline}
where each $A_{ij}$, $B_i$, and $C_i$ is a function only of $(M,\ell,R,\delta M,v_s,\omega)$. The solution is found by multiplying the inverse of the $A_{ij}$ coefficient matrix\footnote{We have no evidence suggesting this matrix is ever non-invertible.} with the right hand side. For brevity, we omit the closed forms for both the coefficients of and solution to Eq. \eqref{eq: polar matrix equation}, as the results are quite lengthy and generally not useful for gaining physical intuition regarding this system. 

The axial version of this procedure admitted a much simpler solution (Eq. \eqref{eq: axial jumps solved}), owing to the lack of coupling between axial perturbations and the motion of the shell.

\section{Perturbative Fundamental Mode Frequency Shift}
\label{app: fundamental mode shift}
\begin{figure}
    \centering
    \includegraphics[width=\linewidth]{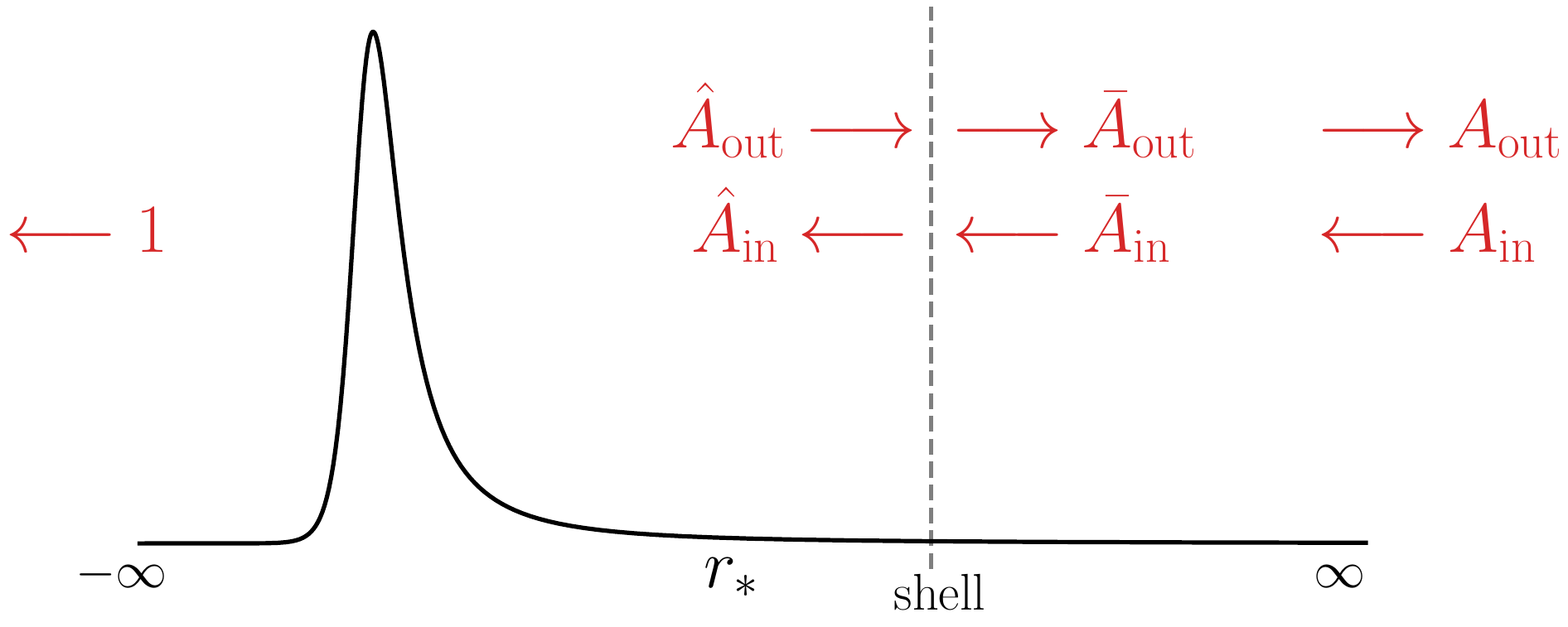}
    \caption{A schematic diagram of the Zerilli potential $V_Z(r_*)$ (black) along with the ``in" solution ingoing and outgoing amplitudes $A_{\text{in}}$ and $A_{\text{out}}$, evaluated just inside the shell (hats), just outside the shell (bars), and at infinity (no marking).}
    \label{fig:Wave Coefficients}
\end{figure}

In this appendix, we provide a more detailed explanation as to why the fundamental polar QNM converges to its unperturbed frequency in the $\delta M\rightarrow 0$ limit, regardless of the shell radius, with a scaling proportional to $\delta M$ (as described in \ref{sec: migrating modes}). Note that while we consider just the polar mode in the following, the same arguments can be applied almost identically to the axial case, thus obtaining the $\delta M$ scaling observed in the main text for this sector as well.

Consider a wave solution which is ingoing at the horizon, and evolve it out to a shell located at a large radius $R$. It is always possible to find coefficients $\hat A_{\text{in}}(\omega,R)$ and $\hat A_{\text{out}}(\omega, R)$ (see Fig. \ref{fig:Wave Coefficients}) such that
\begin{subequations}
    \begin{equation}
        Z^{\text{in}}_{\text{int}}(\omega,R)=\hat A_{\text{in}}(\omega,R)e^{-i\omega R_*^-}+\hat A_{\text{out}}(\omega,R)e^{i\omega R_*^-},
    \end{equation}
    \begin{multline}
        \partial_{r_*}Z^{\text{in}}_{\text{int}}(\omega,r_*)\Bigr|_R=i\omega\Bigr(-\hat A_{\text{in}}(\omega,R)e^{-i\omega R_*^-}
        \\
        +\hat A_{\text{out}}(\omega,R)e^{i\omega R_*^-}\Bigr).
    \end{multline}
\end{subequations}
It follows that
\begin{equation}
    \hat A_{\text{in}}(\omega,R)=\frac{e^{i\omega R_*^-}}{2}\Bigr(Z^{\text{in}}(\omega,R)+\frac{i}{\omega}\partial_{r_*}Z^{\text{in}}(\omega,r_*)|_{R}\Bigr).
    \label{eq: Ain hat}
\end{equation}
To find the coefficients $\bar A_{\text{in}}$ and $\bar A_{\text{out}}$ just outside the shell, we must apply the junction conditions and account for the shift in frequency due to the mass of the shell. However, the junction conditions are far more mathematically complicated than the frequency shift; for now, we neglect the explicit junction conditions so that $Z^{\text{in}}_{\text{int}}(R)=Z^{\text{in}}_{\text{ext}}(R)$ and ${Z^{\text{in}}_{\text{int}}}'(R)={Z^{\text{in}}_{\text{ext}}}'(R)$. We will see later that reintroducing the junction conditions does not alter our conclusions. 

Since the $r_*$ derivative differs by a factor of $\alpha^{-2}$ from the interior to the exterior region, after accounting for the reduction in frequency by $\alpha$, we find
\begin{equation}
    \bar A_{\text{in}}(\omega,R)=\frac{e^{i(\omega/\alpha) R_*^+}}{2}\Bigr(Z^{\text{in}}(\omega,R)+\frac{i}{\omega\alpha}\partial_{r_*}Z^{\text{in}}(\omega,r_*)|_{R}\Bigr).
\end{equation}
In the case where $R\gg M$, the potential is nearly zero in the entire exterior region. Therefore, when the solution is evolved in the exterior regime, using $\bar A_{\text{in}}$ and $\bar A_{\text{out}}$ to produce the initial conditions, the ingoing and outgoing pieces will not substantially mix. Therefore, $A_{\text{in}}(\omega)\approx \bar A_{\text{in}}(\omega, R)$ and $A_{\text{out}}(\omega)\approx\bar A_{\text{out}}(\omega, R)$.

We are interested in finding the perturbed frequency $\omega_0+\delta\omega$ at which $A_{\text{in}}=0$. To do so, we expand around the unperturbed frequency, writing 
\begin{multline}
    A_{\text{in}}(\omega_0+\delta\omega)\propto \Bigr(Z^{\text{in}}+\delta\omega\frac{\partial}{\partial\omega}Z^{\text{in}}\Bigr)
    \\
    +\frac{i}{\alpha(\omega_0+\delta\omega)}\Bigr(\partial_{r_*}Z^{\text{in}}+\delta\omega\frac{\partial}{\partial\omega}(\partial_{r_*}Z^{\text{in}})\Bigr),
\end{multline}
with all instances of $Z^{\text{in}}$ and its derivatives evaluated at $r=R$ and $\omega=\omega_0$. Since $\omega_0$ is an unperturbed QNM frequency, by the time the wave reaches the distant shell (but before it propagates through the shell), it should be an almost entirely outgoing solution, so $\hat A_{\text{in}}(\omega_0,R)\approx 0$. Continuing the expansion to include $\delta M/M\ll1$ and $R\gg M$, and using Eq. \eqref{eq: Ain hat} to eliminate a combination of terms equal to $\hat A_{\text{in}}(\omega_0,R)$, we find
\begin{multline}
    A_{\text{in}}(\omega_0+\delta\omega)\propto\delta\omega\frac{\partial}{\partial\omega}Z^{\text{in}}(\omega,R)\Bigr|_{\omega_0}
    \\
    +\frac{i}{\omega_0}\delta\omega\frac{\partial}{\partial\omega}(\partial_{r_*}Z^{\text{in}}(\omega,r_*)|_R)\Bigr|_{\omega_0}
    \\
    -\frac{i}{\omega_0}\Bigr(\frac{\delta\omega}{\omega_0}+\frac{\delta M}{R-2M}\Bigr)\partial_{r_*}Z^{\text{in}}(\omega_0,r_*)|_R.
    \label{eq: Ain expansion}
\end{multline}
Once again writing $Z^{\text{in}}$ at the large radius shell as the sum of ingoing and outgoing waves, where at $\omega_0$, $Z^{\text{in}}(\omega,R)\sim e^{i\omega R_*^-}$, we approximate
\begin{subequations}
    \begin{multline}
        \frac{\partial}{\partial\omega}Z^{\text{in}}(\omega,R)\Bigr|_{\omega_0}\approx (iR_*^-\hat A_{\text{out}}(\omega_0)+\partial_\omega \hat A_{\text{out}}(\omega)|_{\omega_0})e^{i\omega_0R_*^-}
        \\+\partial_\omega\hat A_{\text{in}}(\omega)|_{\omega_0}e^{-i\omega_0R_*^-},
    \end{multline}
    \begin{equation}
        \partial_{r_*}Z^{\text{in}}(\omega_0,r_*)|_R\approx i\omega_0\hat A_{\text{out}}(\omega_0)e^{i\omega_0R_*^-},
        \label{eq: partial rstar Zin}
    \end{equation}
    \begin{multline}
        \frac{\partial}{\partial\omega}(\partial_{r_*}Z^{\text{in}}(\omega,r_*)|_R)\Bigr|_{\omega_0}\approx i\hat A_{\text{out}}(\omega_0)e^{i\omega_0R^-_*}
        \\
        +i\omega_0(iR_*^-\hat A_{\text{out}}(\omega_0)+\partial_\omega \hat A_{\text{out}}(\omega)|_{\omega_0})e^{i\omega_0 R_*^-}
        \\
        -i\omega_0\partial_\omega \hat A_{\text{in}}(\omega)|_{\omega_0}e^{-i\omega_0 R_*^-},
    \end{multline}
\end{subequations}
where in Eq. \eqref{eq: partial rstar Zin} we approximate the wave amplitude coefficients as independent of $r_*$ because for large shell radii, $V_{\text{Z}}(R)\ll \omega^2$.

With these results, and after some simplification, setting Eq. \eqref{eq: Ain expansion} to zero returns
\begin{equation}
    2\delta\omega\partial_\omega\hat A_{\text{in}}(\omega)|_{\omega_0}e^{-i\omega_0R_*^-}
    \\
    +\frac{\delta M}{R-2M}\hat A_{\text{out}}(\omega_0)e^{i\omega_0R_*^-}\approx 0
    \label{Ain zero condition expanded}
\end{equation}
or
\begin{equation}
    \delta\omega=-\frac{\delta M}{2(R-2M)}\frac{\hat A_{\text{out}}(\omega_0)}{\partial_\omega \hat A_{\text{in}}(\omega)|_{\omega_0}}e^{2i\omega_0R_*^-}.
\end{equation}
Since the solutions which generate $\hat A_{\text{in}}$ and $\hat A_{\text{out}}$ are stopped before they reach the shell, these coefficients are independent of $\delta M$, and thus $\delta \omega\propto\delta M$.

Revisiting the junction conditions, we have checked numerically that in the limit $\delta M/M\ll 1$, their effect on $Z^{\text{in}}$ and $\partial_{r_*}Z^{\text{in}}$ in the exterior region is also linear in $\delta M$, just like the reduction in frequency by $\alpha$ (as seen in Eq. \eqref{eq: Ain expansion}). So, with the junction conditions included, the conclusion that $\delta\omega\propto\delta M$ still holds.

When the shell is located at a smaller radius, we no longer have $A_{\text{in}}(\omega)\approx\bar A_{\text{in}}(\omega,R)$ and $A_{\text{out}}(\omega)\approx\bar A_{\text{out}}(\omega,R)$. However, because the Zerilli equation is a linear differential equation, we can still write the solution at infinity as the sum of the unperturbed solution and shell-driven perturbation ($\propto\delta M$) computed outside the shell and then integrated to infinity. The unperturbed solution at a frequency $\omega_0+\delta \omega$ integrated out to infinity will not only produce a contribution to $A_{\text{in}}$ which again scales with $\delta\omega$, due to the linear expansion around $\omega_0$, but will also produce an additional small contribution proportional to $\delta M$, which emerges when the Zerilli equation is integrated through the remainder of the curvature potential outside the shell with the redshifted frequency $\omega/\alpha\approx \omega(1-\delta M/(R-2M))$. Meanwhile, when the shell-driven perturbation is integrated to infinity, it will still appear in $A_{\text{in}}$ as a component $\propto\delta M$, as corrections due to the redshifted frequency will be at least $\mathcal{O}(\delta M\delta \omega)$. Ultimately, the expression for $A_{\text{in}}$ only changes relative to Eq. \eqref{Ain zero condition expanded} by the addition of terms proportional to $\delta M$, and thus the scaling of $\delta \omega$ is unaffected.

\section{Real-Frequency Poles of Gauge Variable $z$}
\label{app: z pole frequencies}
In Sec. \ref{sec: weakly damped mode physics}, we stated that in the limit of $\delta M\rightarrow 0$, the gauge variable $z$ acquires poles at real frequencies which are the roots of a quadratic polynomial in $\omega^2$. For completeness, we now provide the full expression for these frequencies:
\begin{multline}
    2R^2\omega^2=
    \pm\Bigr\{\Bigr(2v_\text{s}^2(2x-1)(3x-\lambda-3)+x(3x-\lambda-4)\Bigr)^2
    \\
    -4x(\lambda+1)\Bigr(3x^2(1+4v_\text{s}^2)-2v_\text{s}^2\lambda+x(v_\text{s}^2(4\lambda-6)-2)\Bigr)\Bigr\}^{1/2}
    \\
    +(2\lambda+6)v_\text{s}^2-(4+18v_\text{s}^2+\lambda+4\lambda v_\text{s}^2)x+(12v_\text{s}^2+3)x^2,
\end{multline}
where $x\equiv M/R_\text{shell}$. Our numerical tests have indicated that for physical system parameters ($R>2M$, $\ell>0$, $v_s^2<1$), a positive solution for $\omega^2$ emerges if and only if the plus sign is taken, ensuring the appearance of exactly one resonant frequency on the (positive) real axis.

\section{Radial Infall Waveform}
\label{app: radial infall}
Here, we briefly outline the procedure for generating the gravitational waveform for a radial infall trajectory starting from a finite radius (rather than the traditional example of starting at infinity). When the particle starts inside the shell radius, we can take advantage of the fact that the metric is the standard Schwarzschild form with mass $M$ along the entire trajectory.

The existence of the Killing vector $\xi^\alpha=(1,0,0,0)$ in the interior spacetime admits the conserved quantity
\begin{equation}
    e=\Bigr(1-\frac{2M}{r}\Bigr)\frac{dt}{d\tau}.
\end{equation}
Since the particle starts at rest, its initial four velocity $u_{\text{init}}$ must be proportional to $\xi$, and the normalization $u_\alpha u^\alpha=-1$ ultimately leads to 
\begin{equation}
    e=\sqrt{1-\frac{2M}{r_0}},
\end{equation}
where $r_0$ is the initial radial coordinate. Following the standard procedure for computing the radial motion of the particle (such as in \cite{HartleTextbook}), we find
\begin{equation}
    -\frac{M}{r_0}=\frac12\Bigr(\frac{dr}{d\tau}\Bigr)^2+V_{\text{eff}}(r),
\end{equation}
where for an orbit with no angular momentum, $V_{\text{eff}}(r)=-M/r$. We can write the previous equations as
\begin{subequations}
    \begin{equation}
        \frac{dr}{d\tau}(r)=-\sqrt{2M\Bigr(\frac{1}{r}-\frac{1}{r_0}\Bigr)},
    \end{equation}
    \begin{equation}
        \frac{dt}{d\tau}(r)=\frac{\sqrt{1-2M/r_0}}{1-2M/r},
    \end{equation}
\end{subequations}
and combining these equations produces
\begin{equation}
    \frac{dr}{dt}(r)=-\sqrt{2M\Bigr(\frac{1}{r}-\frac{1}{r_0}\Bigr)}\frac{1-2M/r}{\sqrt{1-2M/r_0}},
\end{equation}
which can be integrated to find $t(r)$. This trajectory of course is only valid for $r\leq r_0$.

With this in hand, we now consider the gravitational radiation waveform. As discussed in \cite{MaggioreVol2}, the source term for the Zerilli equation (normalized by the mass of the infalling particle) is
\begin{multline}
    S_\ell(\omega,r)=-4\sqrt{2\pi(\ell+1/2)}\frac{1-2M/r}{\lambda r+3M}
    \\
    \times\Bigr[\sqrt{\frac{r}{2M}}-\frac{2i\lambda}{\omega(\lambda r+3M)}\Bigr]e^{i\omega t(r)}.
\end{multline}
We suppress the azimuthal index $m$ in $S_{\ell}$ because the source term vanishes for all $m\not=0$. The sourced Zerilli function far from the black hole and shell is then
\begin{equation}
    \mathcal{Z}_\ell(\omega,r_*)=\int_{-\infty}^{r_{*0}}dr'_*G(\omega,r_*,r_*')S_\ell(\omega,r(r_*')),
\end{equation}
where $G_\ell(\omega,r_*,r_*')=Z_\ell^{\text{in}}(\omega, r_*')Z_\ell^{\text{up}}(\omega, r_*)/W_\ell(\omega)$ is the Green's function and $W_\ell(\omega)$ is the Wronskian between the ``in" and ``up" solutions $Z^{\text{in}}(\omega,r_*)$ and $Z^{\text{up}}(\omega,r_*)$ -- recall that $W_\ell(\omega)\propto A_{\text{in},\ell}(\omega)$. The integral is truncated at $r_{*0}\equiv r_*(r_0)$ because the particle does not move through radii outside this point. Since the ``up" solution $Z^{\text{up}}$ behaves like $e^{i(\omega/\alpha) r_*}$ in the far-field regime, after defining \begin{equation}
    \mathcal{A}_\ell(\omega)\equiv\frac{1}{W_\ell(\omega)}\int_{-\infty}^{r_{*0}}dr'_*Z_\ell^{\text{in}}(\omega, r_*')S_\ell(\omega,r(r_*')),
\end{equation}
we finally obtain $\mathcal{Z}(u)$ for the retarded time $u\equiv t-r_*$,
\begin{equation}
    \mathcal{Z}_\ell(u)=\int_{-\infty}^{\infty}d(\omega/\alpha)A_\ell(\omega)e^{-i(\omega/\alpha)u}.
\end{equation}
Note that for our waveform plots, $t-t_{\text{peak}}=u-u_{\text{peak}}$.

\bibliography{main}
\end{document}